\documentclass[12pt]{iopart}

\usepackage{graphicx}
\usepackage{xcolor}
\usepackage{comment}
\usepackage{float}
\usepackage{hyperref}
\usepackage{pdfpages}
\expandafter\let\csname equation*\endcsname\relax

\expandafter\let\csname endequation*\endcsname\relax

\usepackage{amsmath}
\usepackage{setspace}
\usepackage{fancyhdr}
\makeatletter
\def\footnoterule{\kern-3\p@
  \hrule \@width \textwidth \kern 2.6\p@} 
\makeatother
\usepackage[style=numeric-comp,backend=bibtex,sorting=none,isbn=true,url=true,doi=true,maxnames=50]{biblatex}
\begin{document}
\title{Global gyrokinetic analysis of Wendelstein 7-X discharge: unveiling the importance of trapped-electron-mode and electron-temperature-gradient turbulence}

\author{Felix Wilms$^1$, Alejandro Bañón Navarro$^1$, Thomas Windisch$^2$, Sergey Bozhenkov$^2$, Felix Warmer$^3$, Golo Fuchert$^2$, Oliver Ford$^2$, Daihong Zhang$^2$, Torsten Stange$^2$, Frank Jenko$^1$, and the W7-X Team\footnote[7]{See Klinger et al. 2019 (\url{https://doi.org/10.1088/1741-4326/ab03a7}) for the W7-X Team.}}

\address{$^1$Max Planck Institute for Plasma Physics, Boltzmannstr. 2, 85748 Garching, Germany\\
$^2$Max Planck Institute for Plasma Physics, Wendelsteinstr. 1, 17491 Greifswald, Germany\\
$^3$Eindhoven University of Technology, Noord-Brabant, Netherlands}
\ead{{felix.wilms@ipp.mpg.de}}
\vspace{11pt}
\begin{indented}
\item[]February 2024
\end{indented}
\begin{abstract}
We present the first nonlinear, gyrokinetic, radially global simulation of a discharge of the Wendelstein 7-X-like stellarator (W7-X), including kinetic electrons, an equilibrium radial electric field, as well as electromagnetic and collisional effects. By comparison against flux-tube and full-flux-surface simulations, we assess the impact of the equilibrium ExB-flow and flow shear on the stabilisation of turbulence. In contrast to the existing literature, we further provide substantial evidence for the turbulent electron heat flux being driven by trapped-electron-mode (TEM) and electron-temperature-gradient (ETG) turbulence in the core of the plasma. The former manifests as a hybrid together with ion-temperature-gradient (ITG) turbulence and is primarily driven by the finite electron temperature gradient, which has largely been neglected in nonlinear stellarator simulations presented in the existing literature. 
\end{abstract}

%
%
%
%
%
\section{Introduction}

Due to the effectiveness in optimising neoclassical physics proven in \cite{beidler2021demonstration}, plasma turbulence has become the key driver of heat and particle transport in the plasma core of the Wendelstein 7-X (W7-X) stellarator \cite{klinger}. 

The most prominent types of instabilities potentially driving turbulence in W7-X plasmas so far are the ion-temperature-gradient-mode (ITG), electron-temperature-gradient-mode (ETG) and trapped-electron-mode (TEM). In highly electromagnetic regimes of future experiments, instabilities such as kinetic-ballooning-modes (KBM), might become relevant \cite{ksenia,mishchenko2023global}.

Although studies of ETG turbulence in stellarators date back as early as 2002 \cite{jenko2002stellarator}, its impact on transport processes in Wendelstein 7-X remains a subject of ongoing debate. It was argued in \cite{plunketg} that it should be negligible for W7-X, while it was concluded in \cite{weir2021heat} that ETGs could be the primary contributor to anomalous electron heat transport in certain experimental discharges. In contrast, extensive studies have focused on turbulence driven by ITG and TEM. It has been suggested that TEM turbulence is relatively weak in W7-X due to the 'maximum-J' property, where most trapped electrons reside in regions of positive average magnetic field curvature \cite{proll2022turbulence,pellet,valley}. However, it is worth mentioning that many, though not all, of those studies consider TEMs that are primarily driven by a finite density gradient rather than a finite electron temperature gradient. As a result, the most significant contributor to turbulence in the core of W7-X was hypothesised to be ITG, with TEMs only expected to be found in the edge region, if at all \cite{klinger,grulkeitgcore}.

Numerical studies on this subject, typically within the framework of gyrokinetic theory, have faced certain limitations. For instance, while works like \cite{warmer2021impact,pellet,estrada2021radial,zhang2023observation,thienpondt2023prevention} consider experimental scenarios with realistic plasma dynamics, their simulations are limited to flux-tube domains. Conversely, global nonlinear simulations, like those in \cite{gene-3d-EM,mishchenko2022gyrokinetic,singh2022global,mishchenko2023global,navarro2023first,xgc-s}, often rely on analytical profiles or use simplifying assumptions, such as assuming an adiabatic electron model, neglecting collisions or neglecting electromagnetic effects. While each giving valuable insights into the behaviour of the underlying scenarios, the combination of both approaches is still unexplored.

We start lifting these constraints in this work by presenting a global simulation of experimental discharge parameters of Wendelstein 7-X using the GENE-3D code \cite{gene-3d,gene-3d-EM}. The simulation employs a kinetic electron model, accounting for electromagnetic effects from magnetic flutter, collisions, and an equilibrium radial electric field. The only neglected effect is magnetic compression, which is expected to have a minor influence due to the overall low plasma-$\beta$ in the scenario. To the best of the authors' knowledge, this simulation marks a pioneering effort of its kind. We aim to compare the obtained heat flux levels against those from flux-tube (FT) and full-flux-surface (FFS) simulations. This comparison will help assess the impact of the radial electric field and its shear while also expanding the scope of diagnostics available for analysing turbulence characteristics. Moreover, we challenge the prevailing hypothesis regarding trapped electron modes (TEM) being benign in the core of W7-X plasmas \cite{klinger,grulkeitgcore}. Concurrently, we assess the influence of electron-scale turbulence, particularly driven by electron temperature gradient modes (ETG), on the overall electron heat flux. Identifying and differentiating between these various turbulence types is highly significant, considering they demand distinct mitigation strategies \cite{proll2015tem,stroteich2022seeking}.

The rest of this paper is structured as follows: section \ref{sec:Discharge_details} introduces the details of the experimental discharge considered throughout this work. In section \ref{sec:ECRH_domain_comparison}, the heat fluxes predicted by simulations with different computational domains are compared and the effect of the radial electric field on the overall transport is discussed. Subsequently, the predicted heat fluxes are compared against a power balance analysis. The characteristics of ion-scale turbulence in the discharge's core region are investigated in section \ref{sec:ECRH_core_physics}, showing evidence for trapped electron modes contributing significantly to the total transport. In section \ref{sec:ETG_core}, the contribution of ETG turbulence on the electron heat flux is explored, and the reasons for its varying strength at different radial positions are discussed. Finally, in section \ref{sec:power_balance_comments}, we justify the significance of these results in the experimental context, even though they may not align perfectly with power balance.

\section{Details of the discharge}
\label{sec:Discharge_details}
\begin{figure}[!h]
    \centering
    \includegraphics[width=0.8\textwidth]{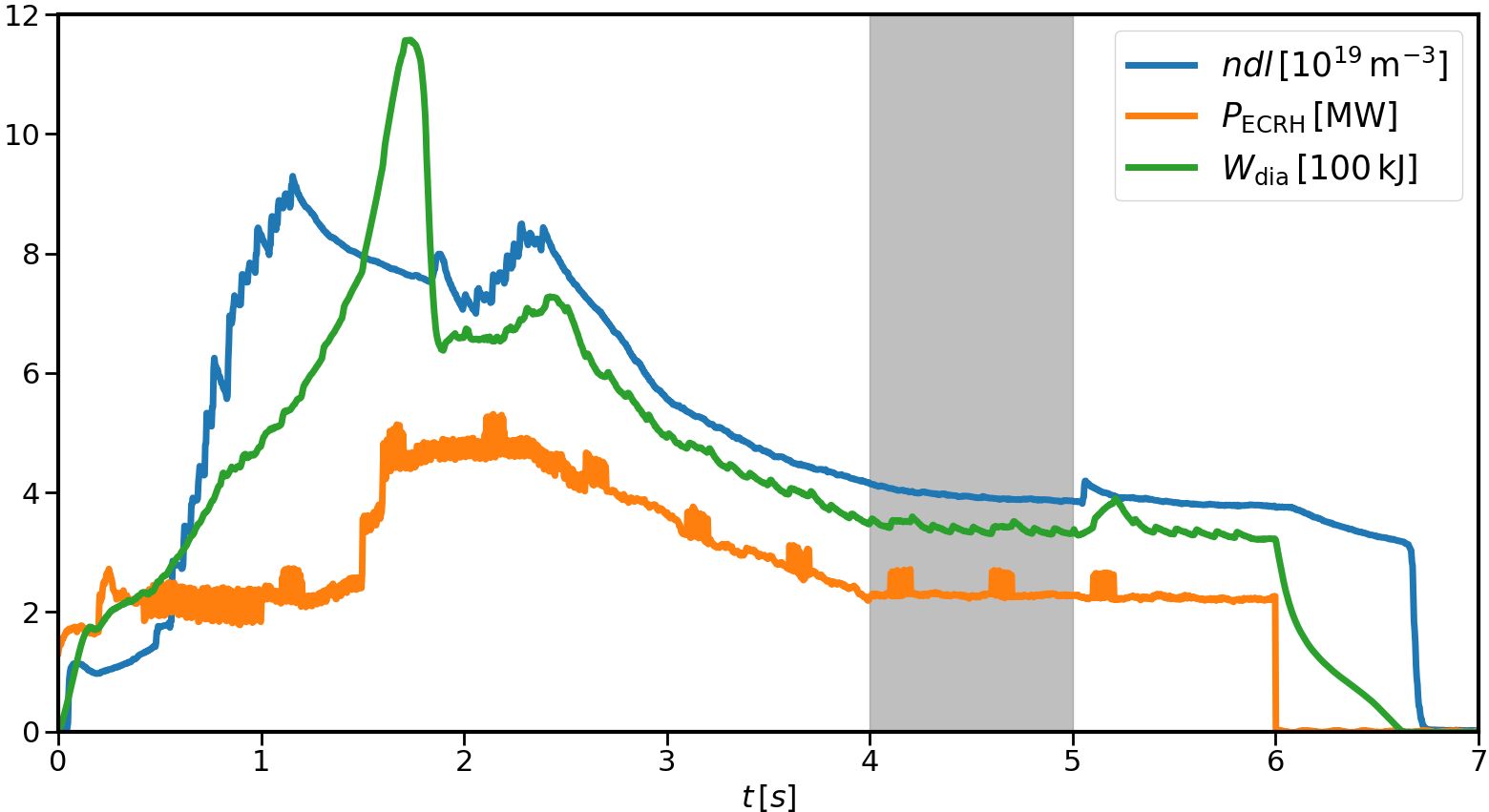}
    \caption{Time traces of the line-integrated density (blue), ECRH heating power (orange) and diamagnetic energy (green) of the W7-X programme 20181016.037. The area marked in grey at $t=4-5\, \mathrm{s}$ corresponds to the shot phase considered in this investigation.}
    \label{fig:ECRH_traces}
\end{figure}

\begin{figure}[!h]
    \centering
    \includegraphics[width=\textwidth]{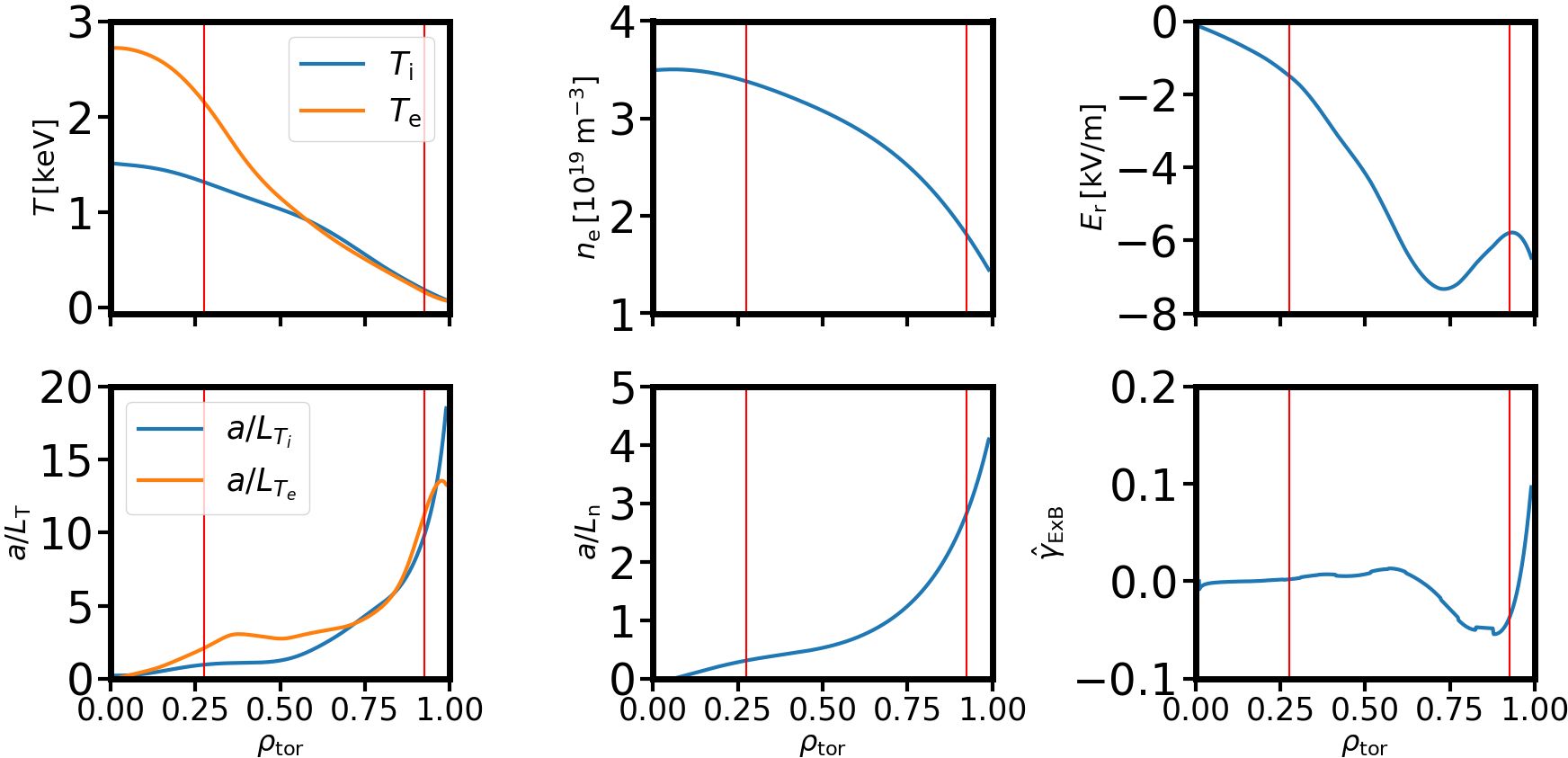}
    \caption{Density, temperature and radial electric field profiles corresponding to the grey area in figure \ref{fig:ECRH_traces}. Top: profiles, bottom: corresponding (normalised) gradients. Bottom right: normalised flow shear rate, defined in equation \ref{eq:ExBrate}. The red lines indicate the limits of the domain simulated with GENE-3D.}
    \label{fig:ECRH_profiles}
\end{figure}

In the rest of this work, we consider an experimental Electron-Cyclotron-Resonance-Heated (ECRH) discharge of Wendelstein 7-X. Specifically, our focus is on the W7-X programme 20181016.037 \cite{pellet}, for which the corresponding time traces can be found in figure \ref{fig:ECRH_traces}.
For our purposes, the discharge phase of $t=4-5\, \mathrm{s}$ is particularly well-suited. It can be considered a representative scenario for gas-fuelled standard discharges within W7-X. The relevant background density and temperature profiles are shown in figure \ref{fig:ECRH_profiles}, together with the equilibrium radial electric field calculated by the neoclassical transport code DKES \cite{dkes} One thing to note is that $T_{\rm i} \geq T_{\rm e}$ for some positions beyond $\rho_{\rm tor}\approx 0.5$. This contradicts the expectations for a plasma solely heated by ECRH, as the ions are only heated through equipartition with the electrons.
However, this discrepancy is of minor concern to us. To analyse turbulent dynamics beyond comparing the different simulation models, we largely focus on the region $\rho_{\rm tor}\leq 0.5$, as will become clear in subsequent discussions.

\section{Comparison of computational domains}
\label{sec:ECRH_domain_comparison}
As a first step, we compare the heat fluxes predicted by flux-tube, full-flux-surface and radially global simulations, as each model offers distinct diagnostic capabilities if satisfactory agreement is reached. For instance, while flux-surface and global simulations can provide the spatial patterns of turbulence across the entire surface, flux-tube simulations can isolate single toroidal modes in linear simulations, which will prove valuable in the subsequent analysis.

The flux-tube simulations presented in this work were performed with the GENE code, whereas full-flux-surface and radially global simulations were performed with GENE-3D. While detailed discussions about their algorithms can be found in \cite{gene,GENE-global} and \cite{gene-3d,gene-3d-EM}, respectively, we note here that both codes use the Clebsh-type spatial coordinates
\begin{equation}
\begin{aligned}
        &x=a\, \rho_{\rm tor}\\
        &y=\sigma_{\rm B_p}\, C_{\rm y}\, \alpha=\sigma_{\rm B_p}\, C_{\rm y}\,(q(x)\theta^*-\phi)\\
        &z=\sigma_{\rm B_p}\, \theta^*.
\end{aligned}
\end{equation}
Here, $\rho_{\rm tor}=\sqrt{\Phi_{\rm tor}\big{/}\Phi_{\rm edge}}$ is used as a radial coordinate, where $\Phi_{\rm tor}$ is the toroidal flux and $\Phi_{\rm edge}$ its value at the last closed flux-surface. The binormal coordinate $y$ is based on the field line label $\alpha$ at a fixed flux-surface, where $q(x)$ is the safety factor, and $\theta^*$ and $\phi$ are the poloidal and toroidal PEST angles \cite{PEST}, respectively. The quantity $\sigma_{\rm B_{\rm p}}$ is the sign of the poloidal magnetic field and ensures that the parallel coordinate $z$ is always in the direction of the background magnetic field $\textbf{B}_0$. Furthermore, the effective minor radius $a=\sqrt{\Phi_{\rm edge}/(\pi B_{\rm ref})}$ is defined via the toroidal magnetic flux at the edge, as well as the magnetic field strength $B_{\rm ref}$ at the axis and the variable $C_{\rm y}$ is defined via the reference position $x_0$ as $C_{\rm y}=x_0/|q(x_0)|$. We use the velocity $v_{||}$ parallel to the magnetic field and the magnetic moment $\mu=m_{\sigma} v_{\perp}^2/(2B_0)$ as velocity space coordinates for each species $\sigma$ . Finally, one can also show that the equilibrium magnetic field $\textbf{B}_0$ can be written as
\begin{equation}
    \textbf{B}_0=B_{\rm ref}\, \hat{\mathcal{C}}(x)\, \nabla x\times \nabla y, 
\end{equation}
where $\hat{\mathcal{C}}= x |q(x_0)|/(x_0 |q(x)|)$.

\begin{figure}
    \centering
    \includegraphics[width=\textwidth]{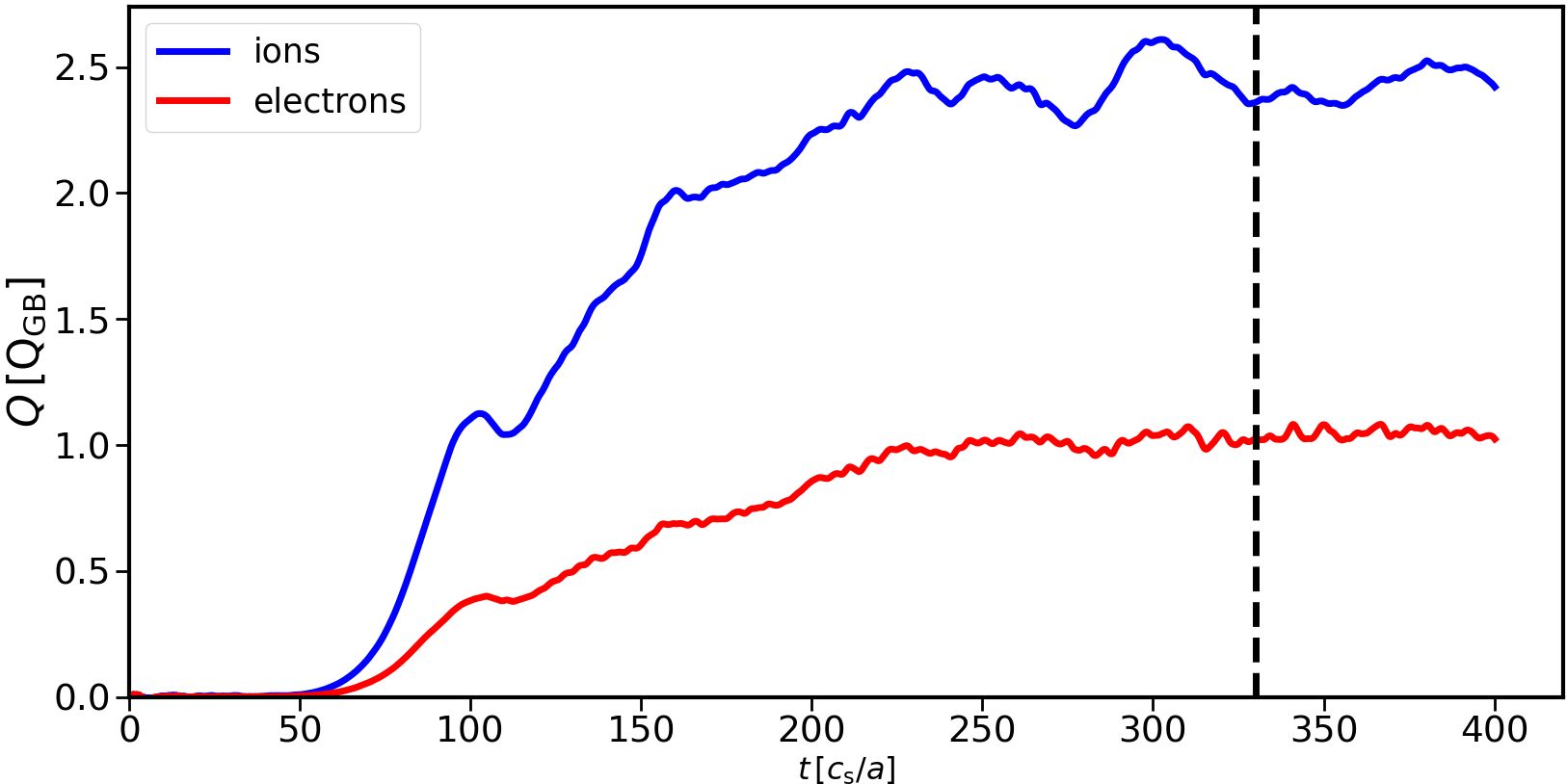}
    \caption{Time trace of the volume-averaged heat flux of the global simulation.}
    \label{fig:trace_global}
\end{figure}

\begin{figure}[h!]
    \centering
    \includegraphics[width=\textwidth]{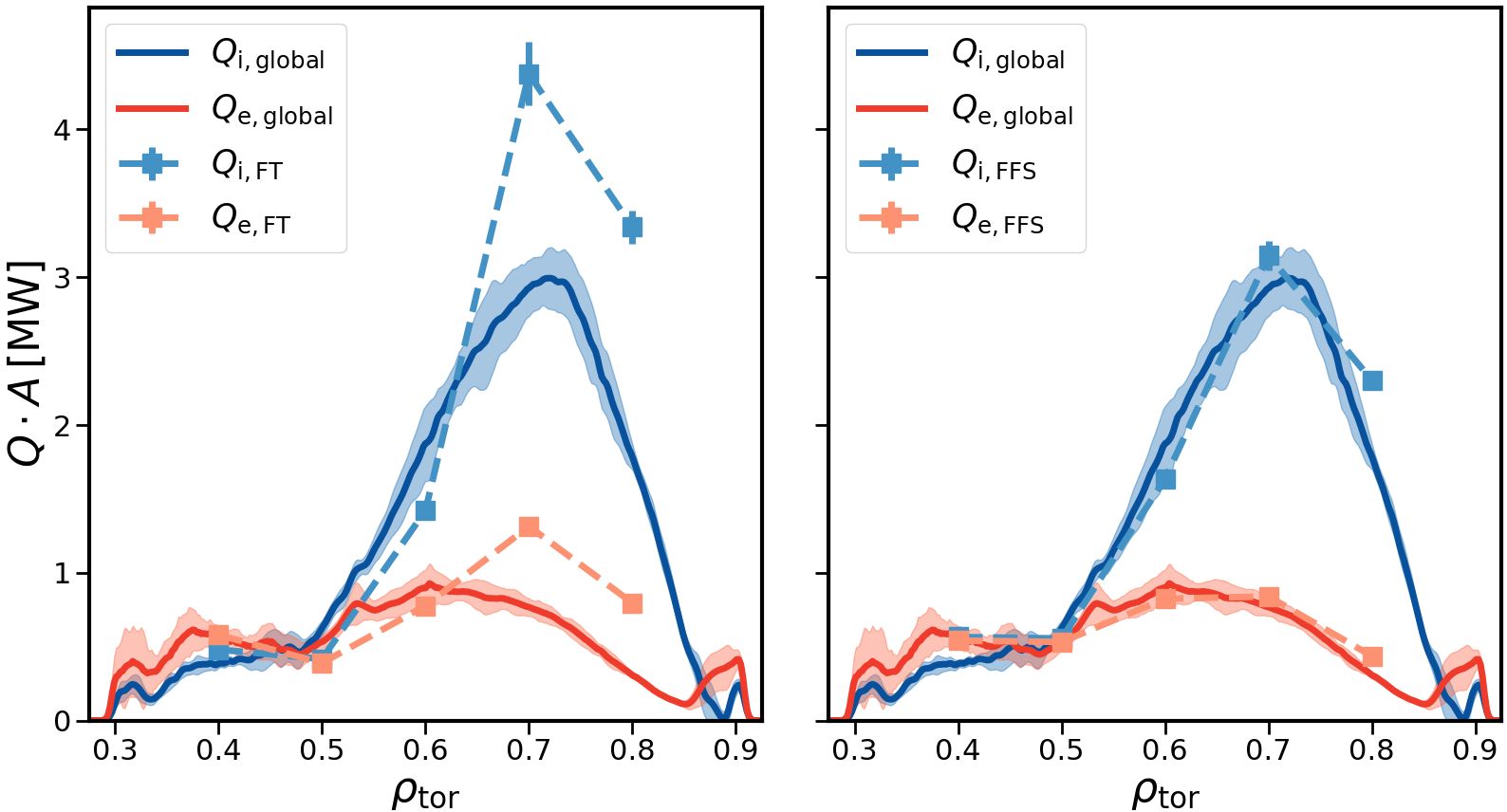}
    \caption{Comparison of the radial heat flux profile obtained from the global simulation with those obtained by flux-tube (left) and full-flux-surface (right) simulations. Shaded regions indicate standard deviation in time of the global simulation.}
    \label{fig:FT_FFS_RG_flux_profiles}
\end{figure}

In the following, we consider the radial domain $\rho_{\rm tor} \in [0.275,0.925]$, which accounts for roughly 65\% of the entire inner plasma volume. The global simulation performed with GENE-3D uses a resolution of $(325,256,126,64,16)$ points in the $(x,y,z,v_{||},\mu)$-directions, respectively. The corresponding numerical box dimensions are $(L_{\rm x},L_{\rm y},L_{\rm v_{\rm ||}},L_{\rm \mu})=(259.437,\rho_{\rm s},265.472 ,\rho_{\rm s},3.45, v_{\rm th,\sigma},11.9, T_{\sigma}(x_0)/B_{\rm ref})$,\\covering one fifth of the surface in the binormal direction by utilising the five-fold symmetry of W7-X. The spatial box is given in units of the ion sound Larmor radius $\rho_{\rm s}=c_{\rm s}/\Omega_{\rm p}$, which is defined via the sound velocity $c_{\rm s}=\sqrt{T_{\rm e}(x_0)/m_{\rm p}}$ and the proton Larmor frequency $\Omega_{\rm p}=eB_{\rm ref}/(m_{\rm p} c)$. The parallel velocity box is given in units of the thermal velocity $v_{\rm th,\sigma}=\sqrt{2T_{\sigma}(x_0)/m_{\sigma}}$ of species $\sigma$. Using approximately 2.8 million core hours on the MPCDF cluster Raven, the simulated plasma is saturated based on the time traces of the volume-averaged heat fluxes illustrated in figure \ref{fig:trace_global}. To the author's knowledge, this simulation marks the first instance of a global gyrokinetic simulation conducted with experimental discharge parameters from W7-X, incorporating kinetic electrons and accounting for electromagnetic effects. The heat flux profiles are then compared with those obtained by radially local simulations, carried out at the radial positions $\rho_{\rm tor}=[0.4,0.5,0.6,0.7,0.8]$. The comparison is shown in figure \ref{fig:FT_FFS_RG_flux_profiles}.
The full-flux-surface simulations are also performed, covering one fifth of the respective surface. Numerical boxes of $(L_{\rm x},L_{\rm v_{\rm ||}},L_{\rm \mu})=(225\,\rho_{\rm s},3\, v_{\rm th,\sigma}(x_0),9\, T_{\rm 0,\sigma}(x_0)/B_{\rm ref})$ and resolutions of $(n_{\rm x},n_{\rm y},n_{\rm z},n_{\rm v_{||}},n_{\mu})=(225,256,128,32,9)$ were sufficient for most cases. However, the velocity space had to be adapted to $(L_{\rm v_{\rm ||}},L_{\rm \mu},n_{\rm v_{||}},n_{\mu})=(6,12,64,12)$ for the simulations at $\rho_{\rm tor}=0.4$ and $0.5$. Unless stated otherwise, a finite equilibrium ExB-flow corresponding to the local value of the nominal radial electric field $E_{\rm r}=-d\phi_0(x)/dx$ is included in the simulations.

Four flux-tube simulations were performed at $\alpha=[0,0.25,0.5,0.75]\pi/5$ using GENE instead of GENE-3D at each radial position. Since it uses a Fourier representation in the binormal direction, the numerical boxes were chosen to be $(k_{\rm y,min},L_{\rm v_{\rm ||}},L_{\rm \mu})=(0.05\, \rho_{\rm s}^{-1}, 3\, v_{\rm th,\sigma}(x_0),9\, T_{\rm 0,\sigma}(x_0)/B_{\rm ref})$ with $k_{\rm x,min}=2\pi/L_{\rm x}$ and $k_{\rm y,min}=2\pi/L_{\rm y}$. Resolutions of $(n_{\rm k_{\rm x}},n_{\rm k_{\rm y}},n_{\rm z},n_{\rm v_{||}},n_{\mu})=(128,64,128,32,9)$ were found to be sufficient. The radial box sizes are selected to be approximately $L_{\rm x}\approx 225 \rho_{\rm s}$ and are adjusted according to constraints imposed by the magnetic shear $\hat{s}(x_0)=x_0\, d\ln(q)/dx{|}_{x=x_0}$ \cite{martin2018parallel}. In cases where the shear becomes too small, periodic boundary conditions in the $z$ direction are assumed.

In figure \ref{fig:FT_FFS_RG_flux_profiles}, we observe a reasonable agreement in heat flux levels between the local and global simulations within the core. However, significant discrepancies arise between the flux-tube and global results in the outer region, reaching up to a factor of two difference at $\rho_{\rm tor}=0.8$. On the contrary, full-flux-surface and global simulations match better, displaying a relative error of approximately 30\% at the same position.
\noindent The discrepancies in figure \ref{fig:FT_FFS_RG_flux_profiles} can partly be attributed to how each simulation model handles the external radial electric field. Flux-tube simulations can eliminate a constant radial electric field by coordinate transformation into a reference frame that rotates with the flow velocity, unlike in full-flux-surface simulations, as the binormal variation of the magnetic field breaks rotational invariance. Global simulations naturally account for the radial variation of the electric field, therefore including a sheared flow in the system, which is known to stabilise turbulence. However, local simulations can approximate ExB-flow shear by linearising the normalised flow velocity around the respective flux-surface: 
\begin{equation}
    \hat{v}_{\rm E_0 }=-\frac{\hat{E}_{\rm r}(x)}{\hat{\mathcal{C}}(x)}\approx -\left(\frac{\hat{E}_{\rm r}(x_0)}{\hat{\mathcal{C}}(x_0)} -\hat{\gamma}_{\rm ExB}(x_0)\,(\hat{x}-\hat{x}_0)\right).
    \label{eq:vExB_linearised}
\end{equation}
Here, $\hat{\gamma}_{\rm ExB}$ represents the normalised shearing rate given by 
\begin{equation}
\begin{aligned}
    \hat{\gamma}_{\rm ExB}(x_0)\equiv& - \frac{d}{d\hat{x}} \left(\frac{\hat{E}_{\rm r}(x)}{\hat{\mathcal{C}}(x)}\right)\bigg{|}_{x=x_0}\\=&-\left[ \frac{d\hat{E}_{\rm r}}{d\hat{x}} -\frac{\hat{E}_{\rm r}}{\hat{x}}\left(1-\hat{s}\right)\right]\bigg{|}_{x=x_0},
    \end{aligned}
    \label{eq:ExBrate}
\end{equation}
with $\hat{x}=\rho_{\rm s}\rho_{\rm tor}$ and $\hat{E}_{\rm r}=T_{\rm e}(x_0) E_{\rm r}/(e a)$.

\begin{figure}[h!]
    \centering
    \includegraphics[width=\textwidth]{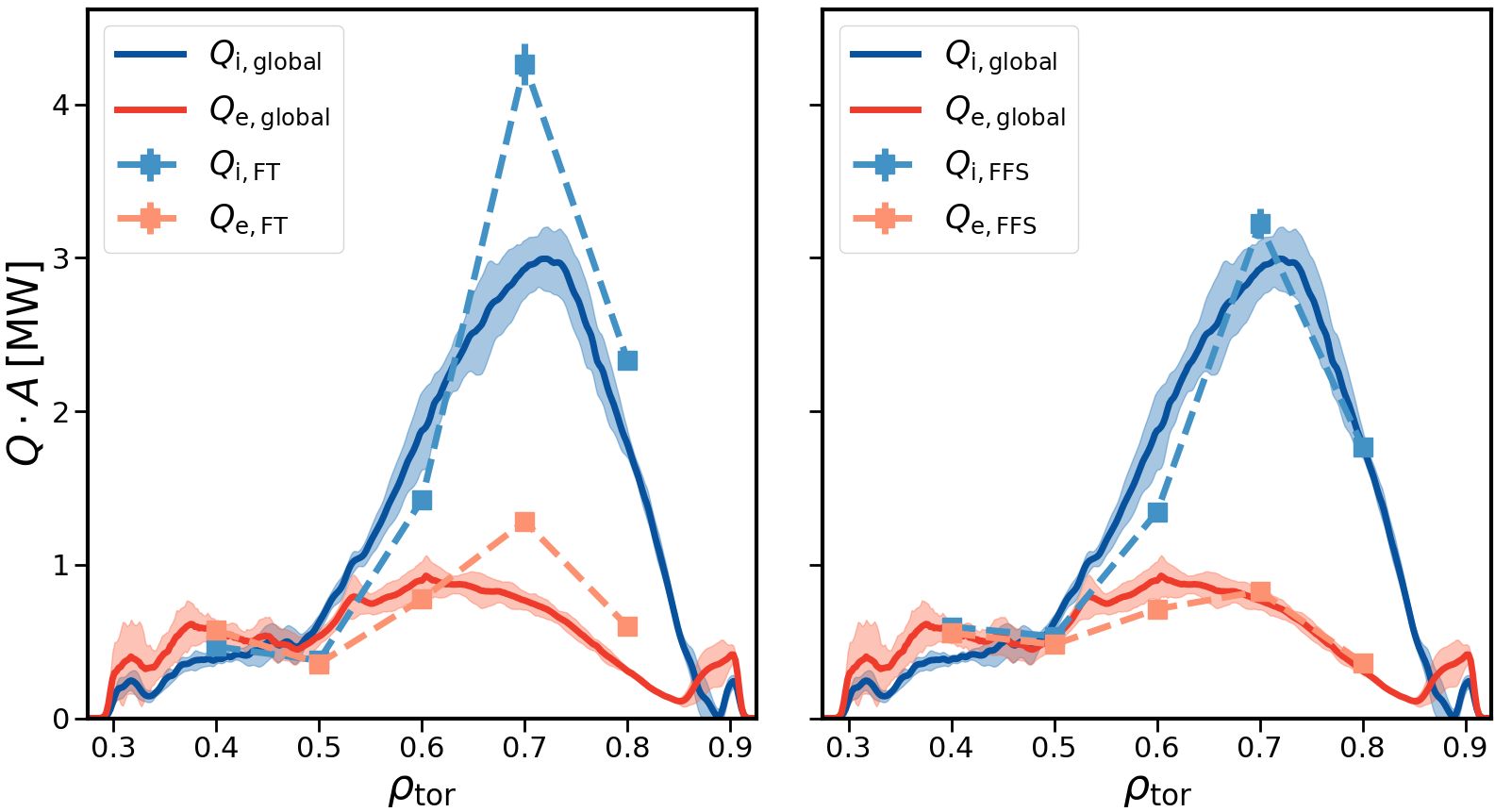}
    \caption{Same as figure \ref{fig:FT_FFS_RG_flux_profiles} but linearised flow shear was added for the radially local simulations.}
    \label{fig:FT_FFS_RG_shear_flux_profiles}
\end{figure}

Incorporating the sheared ExB-flow in the radially local simulations notably influences the outer radial region, as depicted in figure \ref{fig:FT_FFS_RG_shear_flux_profiles}. Given the relatively low shear rate in the inner region, it is understandable that the transport levels remain largely unaffected. 
\begin{table*}
\centering
    \begin{tabular}{c|c|c}
         $ $& $Q_{\rm ions}\cdot A\, \mathrm{[MW]}$& $Q_{\rm electrons}\cdot A\, \mathrm{[MW]}$\\
         \hline
         Flux-tube& $3.34\pm 0.11$ & $0.80\pm 0.02$\\
         \hline
         Flux-tube (with $\hat{\gamma}_{\rm ExB}$)& $2.34\pm 0.04$ & $0.60\pm 0.01$\\
         \hline
         Flux-surface (no $E_{\rm r}$) &$2.32\pm 0.06$&$0.44\pm 0.01$\\
         \hline
         Flux-surface (with $E_{\rm r}$)&$2.30\pm 0.05$&$0.44\pm 0.01$\\
         \hline
         Flux-surface (with $E_{\rm r} \,\&\, \hat{\gamma}_{\rm ExB}$)&$1.77\pm 0.03$&$0.36\pm 0.01$\\
         \hline
         Global &$1.77\pm 0.08$&$ 0.30\pm 0.02$\\
         \hline
    \end{tabular}
    \caption{Ion and electron heat fluxes at $\rho_{\rm tor}=0.8$ as predicted by different models}
    \label{tab:Er_ECRH_comparison}
\end{table*}
\noindent Repeating the full-flux-surface simulation at $\rho_{\rm tor}=0.8$, excluding the constant equilibrium ExB-flow shows that the constant local radial electric field does not notably influence the transport levels at this position, as can be seen in table \ref{tab:Er_ECRH_comparison}. However, when factoring in the flow shear, there is a reduction of about 20-25\% in heat fluxes for both flux-tube and full-flux-surface simulations. With this, the ion heat flux calculated by the latter is reduced to a level agreeing with the global results within error bars. This underscores the significant impact of sheared equilibrium ExB-flows on turbulence in stellarators, emphasising the necessity to consider these dynamics in gyrokinetic simulations. Such considerations are especially crucial in advanced scenarios, like those involving pellet fuelling, where an even more pronounced radial variation in the electric field can be found \cite{pellet}. In addition, we see from figures \ref{fig:FT_FFS_RG_flux_profiles} and \ref{fig:FT_FFS_RG_shear_flux_profiles} that the three simulation models agree well with each other in the core of the plasma, allowing us to use all of them for the subsequent analysis of ion-scale turbulence found in this scenario.

\section{Analysis of ion-scale turbulence in the core}
\label{sec:ECRH_core_physics}
\begin{figure*}[h!]
    \centering
    \includegraphics[width=\textwidth]{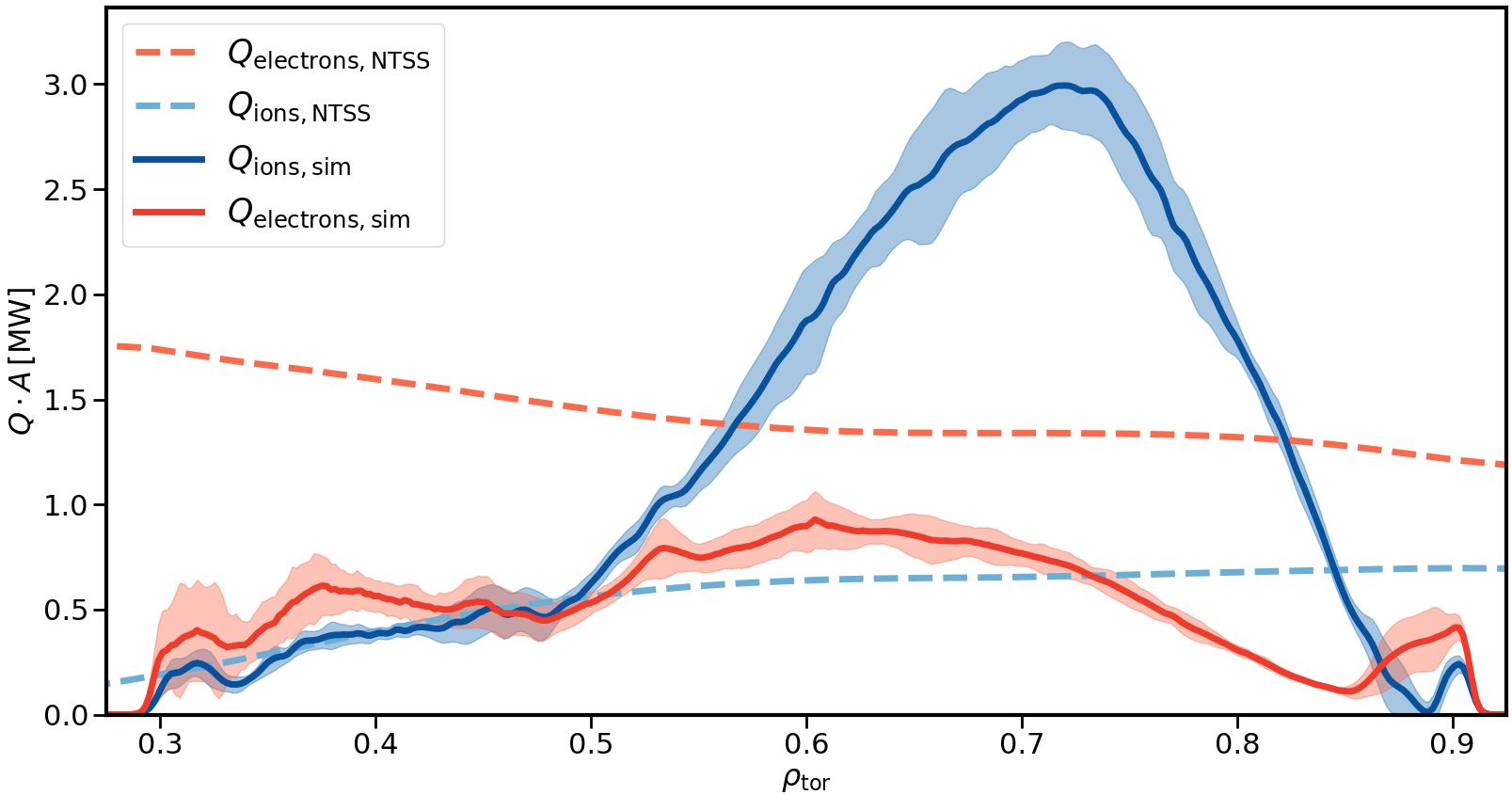}
    \caption{Comparison of the radial heat flux profile obtained from the global simulation with the power balance obtained by NTSS. Shaded regions indicate standard deviation in time of the global simulation.}
    \label{fig:power_balance_ionscale}
\end{figure*}

Having assessed the different simulation models' predictive abilities concerning the current ECRH discharge, we focus on the detailed analysis of its turbulent plasma dynamics. To this end, we compare the global heat flux profiles predicted by GENE-3D against the anomalous heat fluxes calculated with the NTSS code \cite{ntss} as the difference between total and neoclassical heat flux obtained with DKES in figure \ref{fig:power_balance_ionscale}.
The comparison between the power balance and GENE-3D indicates a good agreement in ion heat flux up to $\rho_{\rm tor}=0.5$. Beyond that, GENE-3D predicts heat fluxes well above those given by NTSS, even exceeding the total ECRH heating power shown in figure \ref{fig:ECRH_traces}. In contrast, the electron heat flux is below the power balance prediction over the entire radial domain. Therefore, we primarily consider the region with $\rho_{\rm tor}\leq 0.5$ for in-depth investigations of turbulence characteristics.

It was proposed by \cite{klinger,grulkeitgcore} that ITG would primarily govern the plasma core's turbulence, while any presence of TEMs would likely manifest in the plasma edge. However, the findings depicted in figure \ref{fig:power_balance_ionscale}, both from GENE-3D and NTSS analyses, suggest that this hypothesis might only hold partly true. As supported by \cite{kotschenreuther2019gyrokinetic}, 'pure' ITG turbulence can be expected to drive mainly ion heat flux with only little impact on the electron channel. In contrast, GENE-3D and NTSS predictions predict either $Q_{\rm e}\geq Q_{\rm i}$ or $Q_{\rm e}\gg Q_{\rm i}$ within the core, which can also be anticipated if direct plasma heating is only applied to the electrons. Given that electromagnetic turbulence should be subdominant for the plasma-$\beta$ of the discharge, TEM turbulence emerges as a likely candidate.

\begin{figure}[h!]
    \centering
    \includegraphics[width=\textwidth]{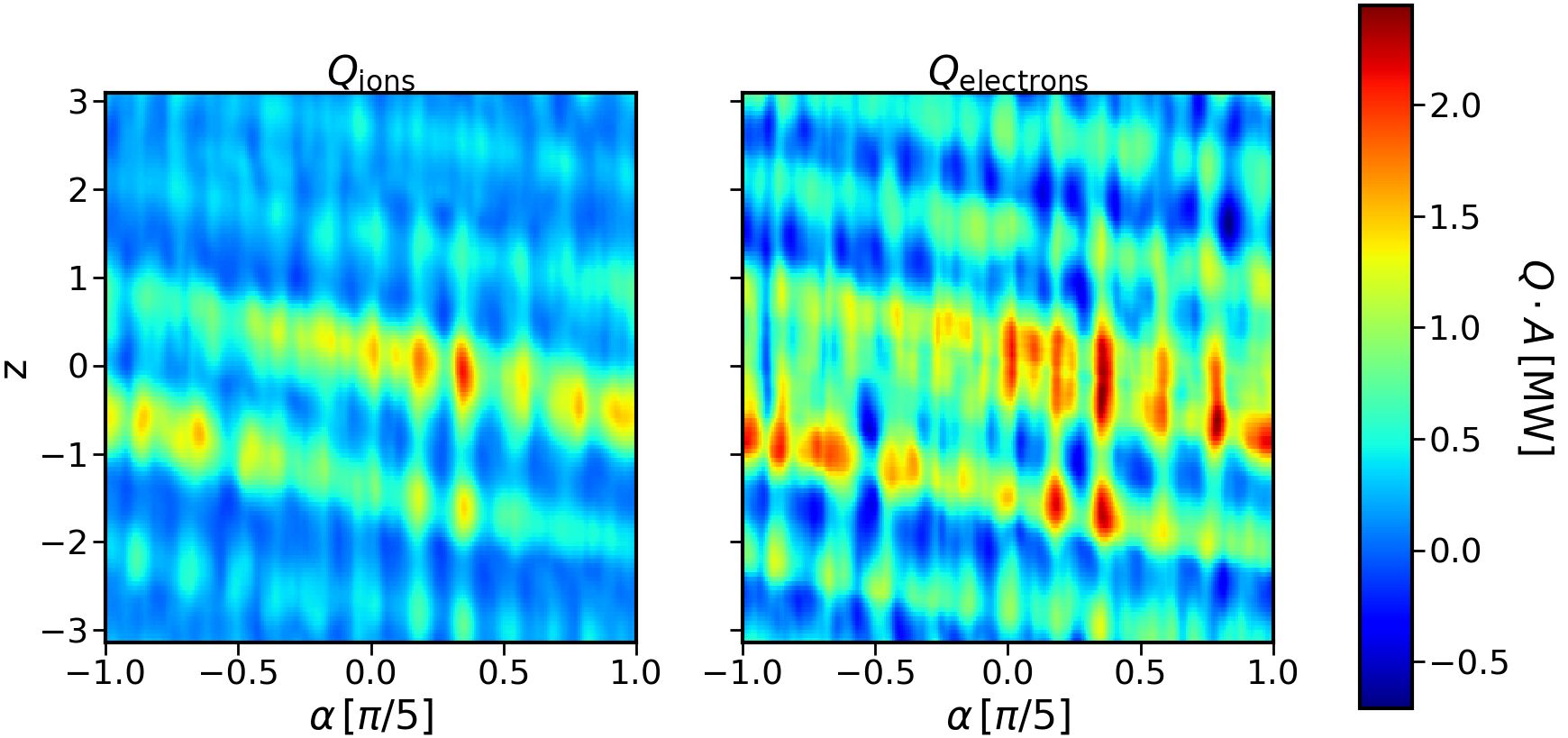}
    \caption{Electron and ion heat fluxes at $\rho_{\rm tor}=0.4$, obtained by the global simulation, as a function of the parallel coordinate $z$ and the field-line label $\alpha$.}
    \label{fig:Q_yz}
\end{figure}

\begin{figure}[h!]
    \centering
    \includegraphics[width=\textwidth]{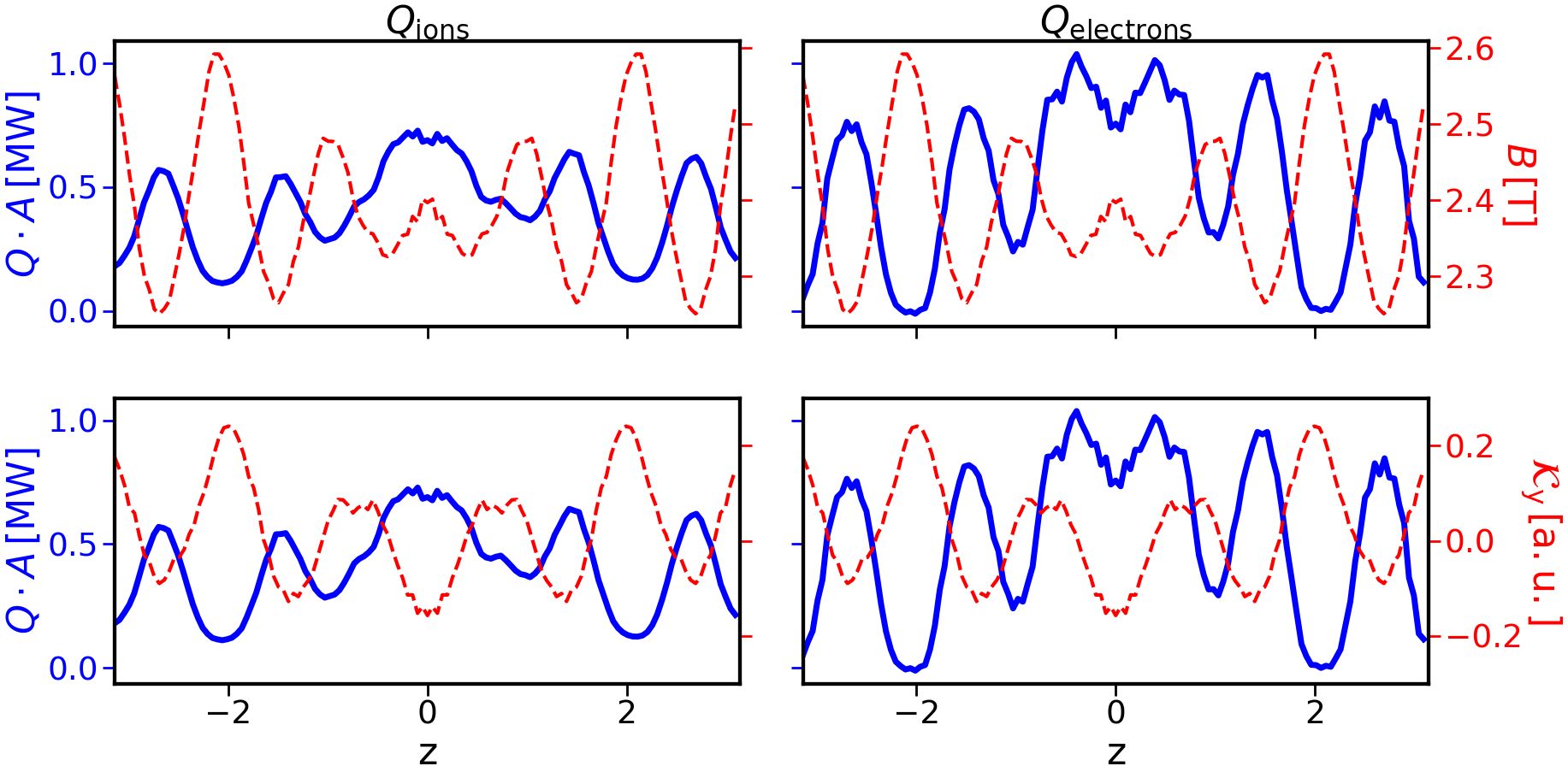}
    \caption{Blue: parallel ion (left) and electron (right) heat flux structures for the flux-tube simulation at $\rho_{\rm tor}=0.4,\, \alpha=0$ with nominal parameters; red: parallel structure of the equilibrium magnetic field strength (top) and the curvature-drive term $\mathcal{K}_{\rm y}$ (bottom).}
    \label{fig:Q_zslice_FT}
\end{figure}

\begin{figure}[h!]
    \centering
    \includegraphics[width=\textwidth]{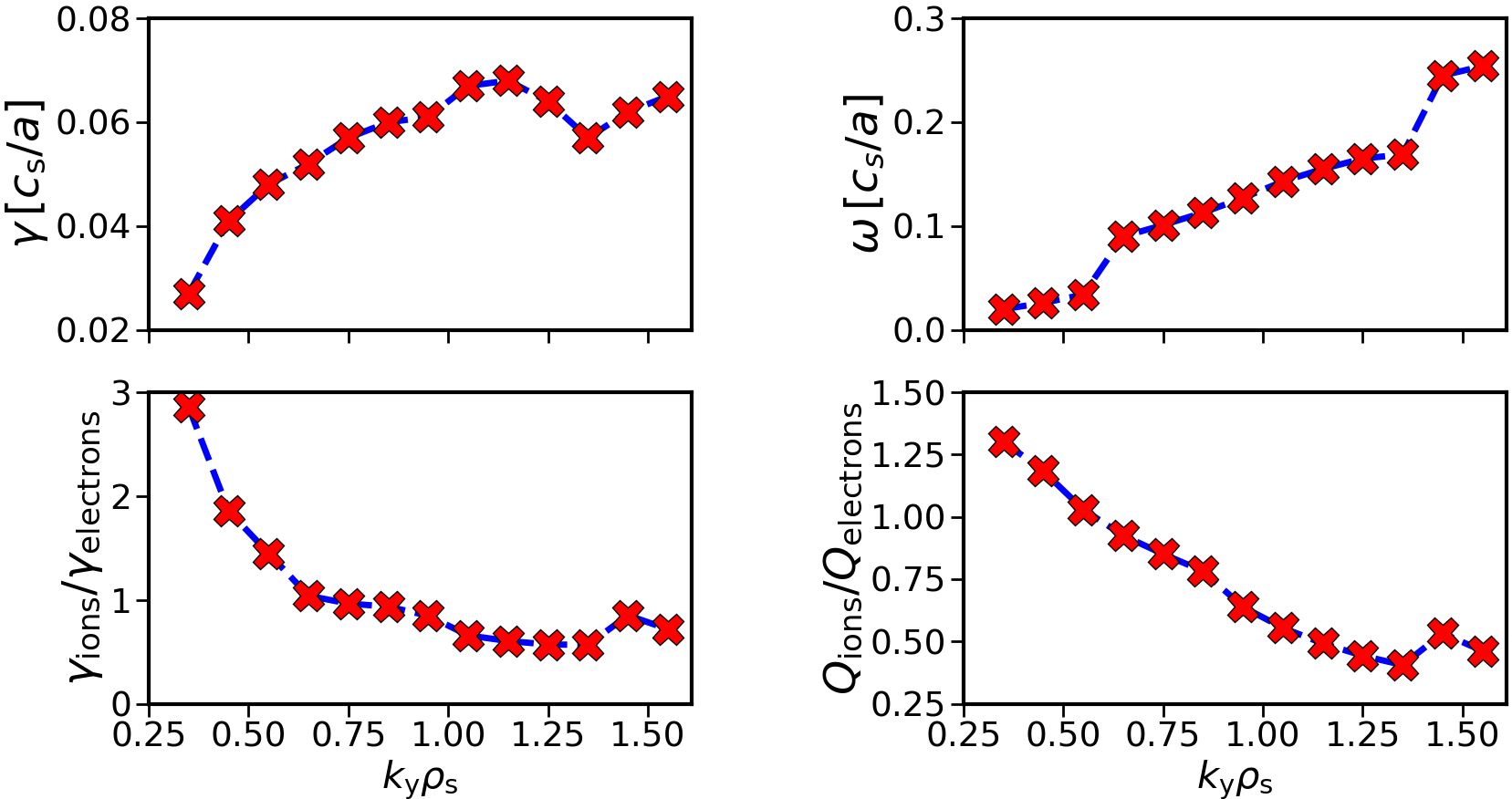}
    \caption{Linear flux-tube wavenumber-scan for the bean-shaped flux-tube at $\rho_{\rm tor}=0.4$; top left: growth rates, top right: mode frequencies, bottom left: ratio of ion and electron contributions to the growth rates, bottom right: quasilinear estimates of the ion-to-electron heat flux ratio.}
    \label{fig:gamma_omega_Ql_x04}
\end{figure}

\begin{figure}[h!]
    \centering
    \includegraphics[width=\textwidth]{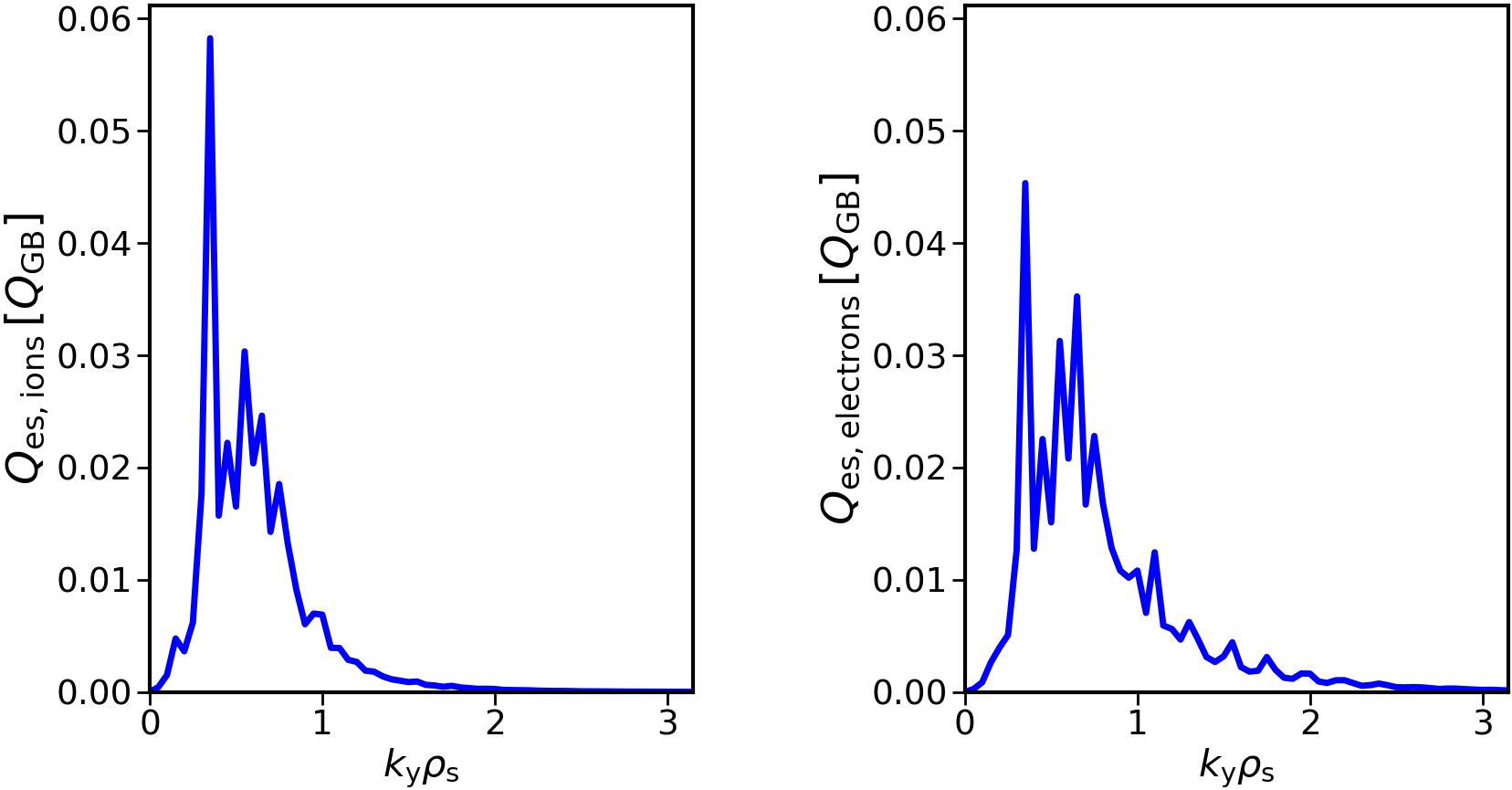}
    \caption{Nonlinear heat flux spectra for the bean-shaped flux-tube at $\rho_{\rm tor}=0.4$; left: ion heat flux, right: electron heat flux.}
    \label{fig:FT_NL_x04}
\end{figure}

We have selected the flux-surface at $\rho_{\rm tor}=0.4$ as representative for detailed analysis within the inner plasma core. Figure \ref{fig:Q_yz} shows the spatial heat flux distribution for electron and ion channels along the parallel coordinate $z$ and the field-line label $\alpha$. Notably, the ion heat flux exhibits a predominant single central peak, contrasting the electron heat flux, which displays multiple maxima along $z$ at fixed $\alpha$.
The distinction becomes notably clearer when comparing the parallel structure at $\alpha=0$ from flux-tube simulations with both the magnetic field structure and the binormal magnetic field curvature $\mathcal{K}_{\rm y}=-\left( \textbf{B}_0\times \nabla |\textbf{B}_0| \right)\cdot \textbf{e}_{\rm y}$ shown in figure \ref{fig:Q_zslice_FT}.
\noindent The ion heat flux seems to peak where $\mathcal{K}_{\rm y}$ is the most negative. Conversely, the electron heat flux shows a more pronounced influence from the magnetic well structure. This is evident as the electron heat flux displays a local minimum at $z=0$, even though the curvature has a local minimum. Furthermore, its local maxima coincide closely with the magnetic wells' minima. These structural variations suggest the presence of a hybrid ITG-TEM mode, further supported by the linear flux-tube simulations depicted in figure \ref{fig:gamma_omega_Ql_x04}.
\noindent In there, all the analysed modes exhibit a positive frequency, therefore propagating in the ion-diamagnetic direction. Nevertheless, an energy transfer analysis \cite{navarro2011free} shows that both ions and electrons are driving the respective instability with comparable contributions. Following the analysis in \cite{collisionless2}, this is indicative of ITG-TEM hybrid modes and rules out dominant ITG and ion-driven trapped-electron-mode (iTEM) \cite{plunk2017collisionless} instabilities, as they are primarily driven by the ions. Additionally, both particle channels have comparable strengths concerning their quasilinear heat flux ratio. This underscores a strong interplay between ions and electrons, serving as an indication of the existence of ITG-TEM hybrids \cite{xanthopoulosquasilinear}. Furthermore, one can notice in the growth rate spectrum that there is no decrease within the scanned wavenumber range. However, it becomes evident that these small-scale modes do not significantly contribute to the nonlinear transport, as the corresponding nonlinear heat flux spectra shown in figure \ref{fig:FT_NL_x04}, are primarily dominated by wavenumbers up to $k_{\rm y}\rho_{\rm s}\approx 1$.

\begin{figure}[h!]
    \centering
    \includegraphics[width=\textwidth]{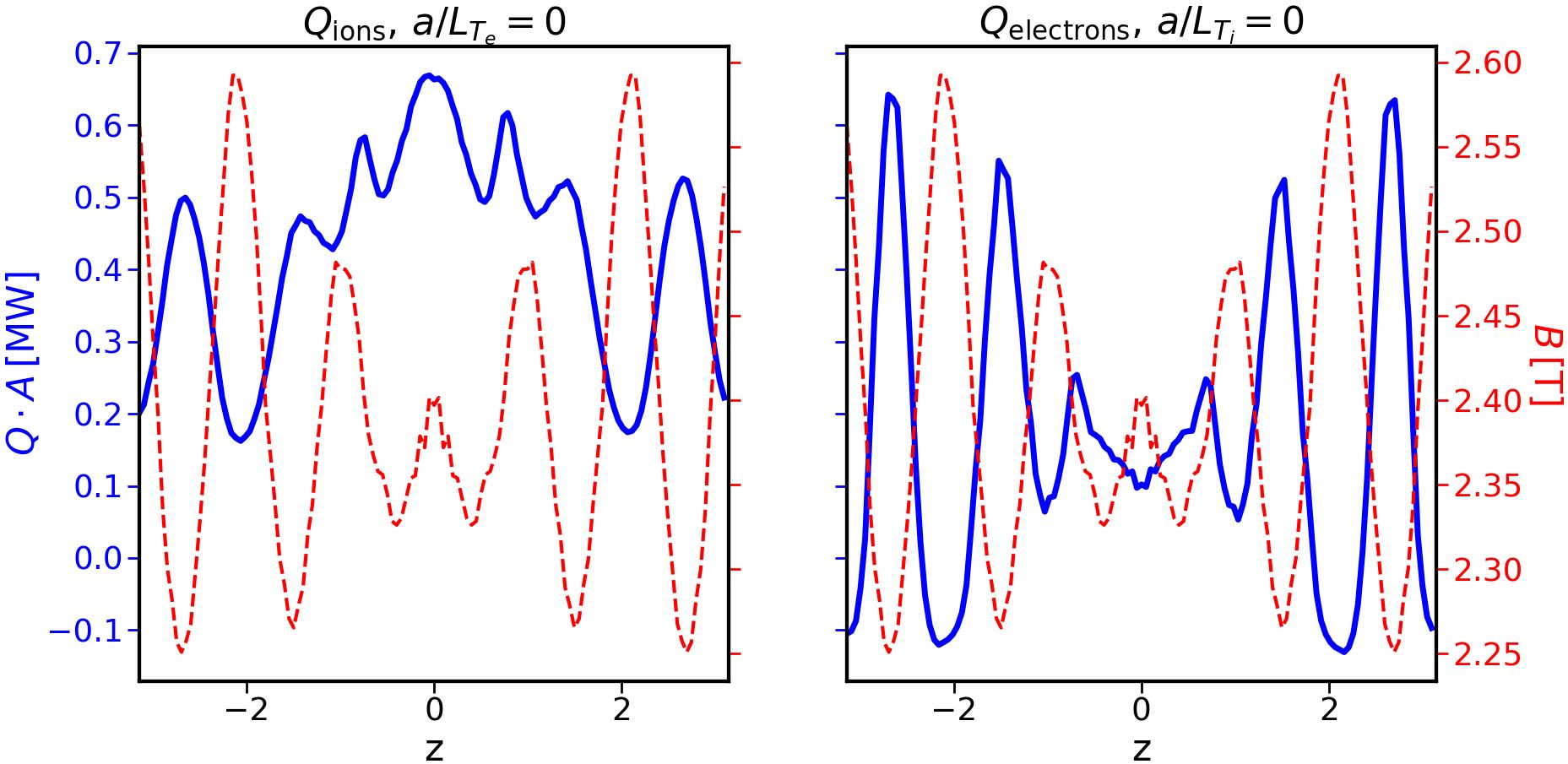}
    \caption{Left: parallel ion heat flux structure for the flux-tube simulation with $a/L_{\rm T_{\rm e}}=0$; right: parallel electron heat flux structure for the flux-tube simulation with $a/L_{\rm T_{\rm i}}=0$. While the heat fluxes are shown in blue, the parallel structure of the magnetic field strength is represented by red dashed lines.}
    \label{fig:slices_omtie}
\end{figure}

We finalise this investigation by assessing the impact of each temperature gradient on the transport channels. While this approach is not fully rigorous, it provides valuable insights into the contributions of various turbulence types to the overall system dynamics.
\begin{table}[h!]
\centering
\begin{tabular}{c|c|c}
 Case& $Q_{\rm ions}\cdot A\, \mathrm{\left[MW\right]}$ &  $Q_{\rm electrons}\cdot A\, \mathrm{\left[MW\right]}$\\
\hline
 Nominal & $0.42\pm 0.05$ & $0.49\pm0.07$ \\
 \hline
 $a/L_{\rm T_{\rm e}}=0$ & $0.49\pm 0.06$ & $0.07\pm 0.01$ \\
\hline
  $a/L_{\rm T_{\rm i}}=0$&$0.01\pm 0.00$&$0.18\pm 0.01$
\end{tabular}
\caption{Heat fluxes of flux-tube simulations at $\rho_{\rm tor}=0.4,\, \alpha=0$, for nominal parameters and either electron or ion temperature gradient set to zero.}
\label{tab:ECRH_FT_contributions}
\end{table}
Table \ref{tab:ECRH_FT_contributions} shows that excluding the normalised electron temperature gradient, compared to the nominal parameters, has a negligible impact on the ion heat flux. It remains similar to the nominal case within error bars. Conversely, in this scenario, the electron heat flux reduces by approximately 85\%, leading to the ion channel dominating the total transport. As shown in figure \ref{fig:slices_omtie}, the modified ion heat flux exhibits a ballooning-type structure along the magnetic field lines, including significantly stronger contributions at the magnetic hills around $z=0$ in contrast to the ion heat flux displayed in figure \ref{fig:Q_zslice_FT}. Comparable structures were documented in \cite{pellet} for a case categorised as ITG-driven, which we conclude is also the dominant type of turbulence.
Furthermore, the ion heat flux vanishes almost completely when the ion temperature gradient is zero. In this scenario, the electron heat flux no longer peaks at $z=0$, where the magnetic field strength reaches a local maximum, as shown in figure \ref{fig:slices_omtie}. Unlike the nominal case presented in figure \ref{fig:Q_zslice_FT}, we thus identify TEM turbulence as the primary driver of transport in this case, drawing its energy from the electron temperature gradient $\nabla T_{\rm e}$. In addition, we see that the electron heat flux experiences a drastic drop by a factor of 7 when the electron temperature gradient is omitted, as indicated in table \ref{tab:ECRH_FT_contributions}. This underlines the significance of this particular branch of trapped electron modes compared to those primarily driven by an electron density gradient, which have received more attention in existing literature \cite{pellet,proll2022turbulence,alcuson2023quantitative,proll2012resilience}.

In summation, evidence derived from the power balance, heat flux structure, and linear flux-tube analysis within this section collectively suggests that in contrast to the proposition outlined in \cite{klinger,grulkeitgcore}, trapped-electron-mode turbulence seems to be present within the core of the gas-fuelled discharge in W7-X, at least for the specific discharge under consideration. However, the applicability of this observation for general experimental scenarios requires further exploration, a task reserved for future investigations.

\section{Impact of ETG turbulence}
\label{sec:ETG_core}
As shown in figure \ref{fig:power_balance_ionscale}, there is a strong alignment between the anticipated anomalous ion transport derived from power balance analysis and the one computed using GENE-3D within the core region. However, the predicted electron transport remains notably lower. It is crucial to notice that our focus has predominantly been on turbulence at ion lengthscales thus far. Consequently, the remaining contribution to the electron flux might be driven by ETG turbulence, which has not yet been captured in our simulations.

To this end, we perform additional flux-tube simulations using an adiabatic ion model. We consider the flux-tubes at $\alpha=[0,0.25,0.5,0.75]\pi/5$ for $\rho_{\rm tor}=[0.3,0.4,0.5,0.6,0.7,0.8,0.9]$. The numerical resolution and box sizes remain the same as those used in the ion-scale simulations in section \ref{sec:ECRH_domain_comparison}, with the exception being the use of the electron sound Larmor radius $\rho_{\rm e}$ for normalisation instead of the ion sound Larmor radius $\rho_{\rm s}$.

\begin{figure}[h!]
    \centering
    \includegraphics[width=\textwidth]{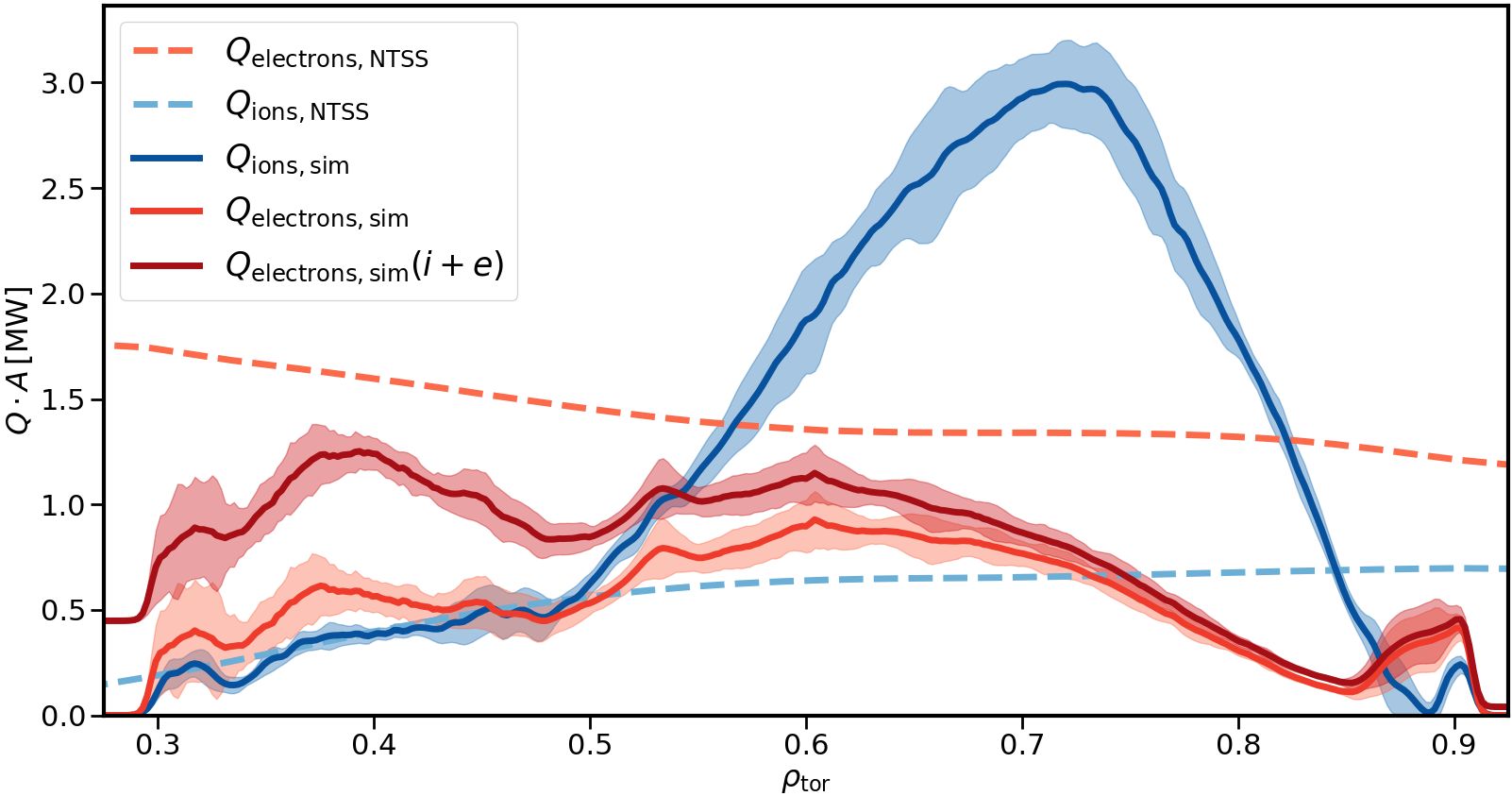}
    \caption{Comparison of the radial heat flux profile with the power balance obtained by NTSS. The dark red line shows the sum of electron heat fluxes of (separated) ion- and electron-scale simulations. Shaded regions indicate standard deviation in time of the global simulation.}
    \label{fig:power_balance_electronscale}
\end{figure}

The resulting fluxes, measured in megawatts, are averaged over the field-line labels and added to the ion-scale electron heat flux predicted by GENE-3D via linear interpolation in the radial coordinate. Figure \ref{fig:power_balance_electronscale} shows the updated global flux profiles.
\noindent In there, we observe that the contribution from electron-scale turbulence is negligible in the outer radial region. Nevertheless, ETG turbulence seems to be responsible for more than 50\% of the turbulent electron heat flux within the plasma core. Our findings corroborate those in \cite{weir2021heat}, which identified ETG turbulence as a possible primary driver for electron heat transport in the core for certain experimental scenarios. However, the electron-scale contribution to their discharges was identified as even stronger than for the present case.

\begin{figure}
    \centering
    \includegraphics[width=\textwidth]{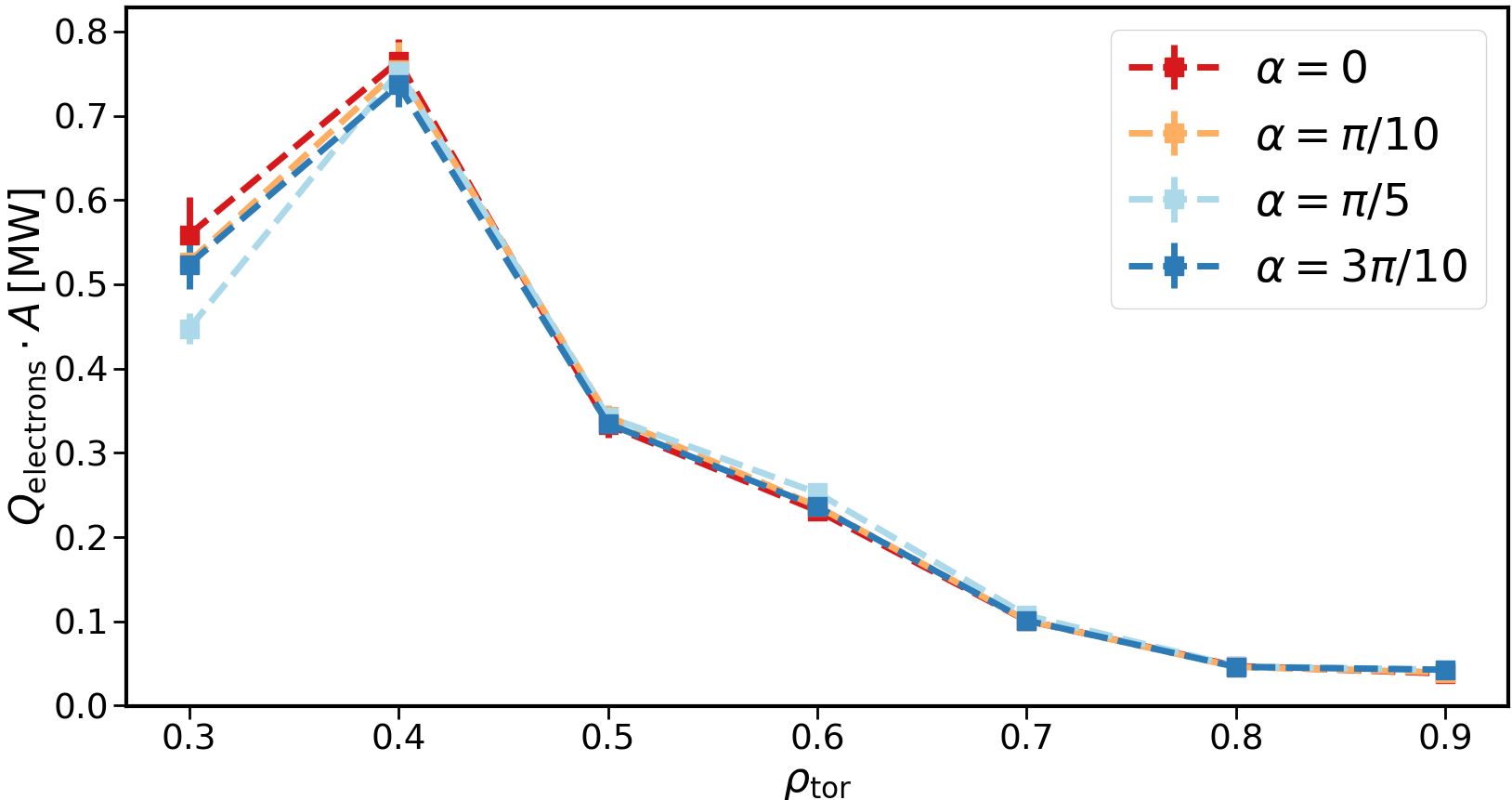}
    \caption{Electron heat fluxes obtained by local ETG simulations as a function of the radial coordinate for different field-line labels.}
    \label{fig:ETG_alpha}
\end{figure}

\begin{figure}[h!]
    \centering
    \includegraphics[width=\textwidth]{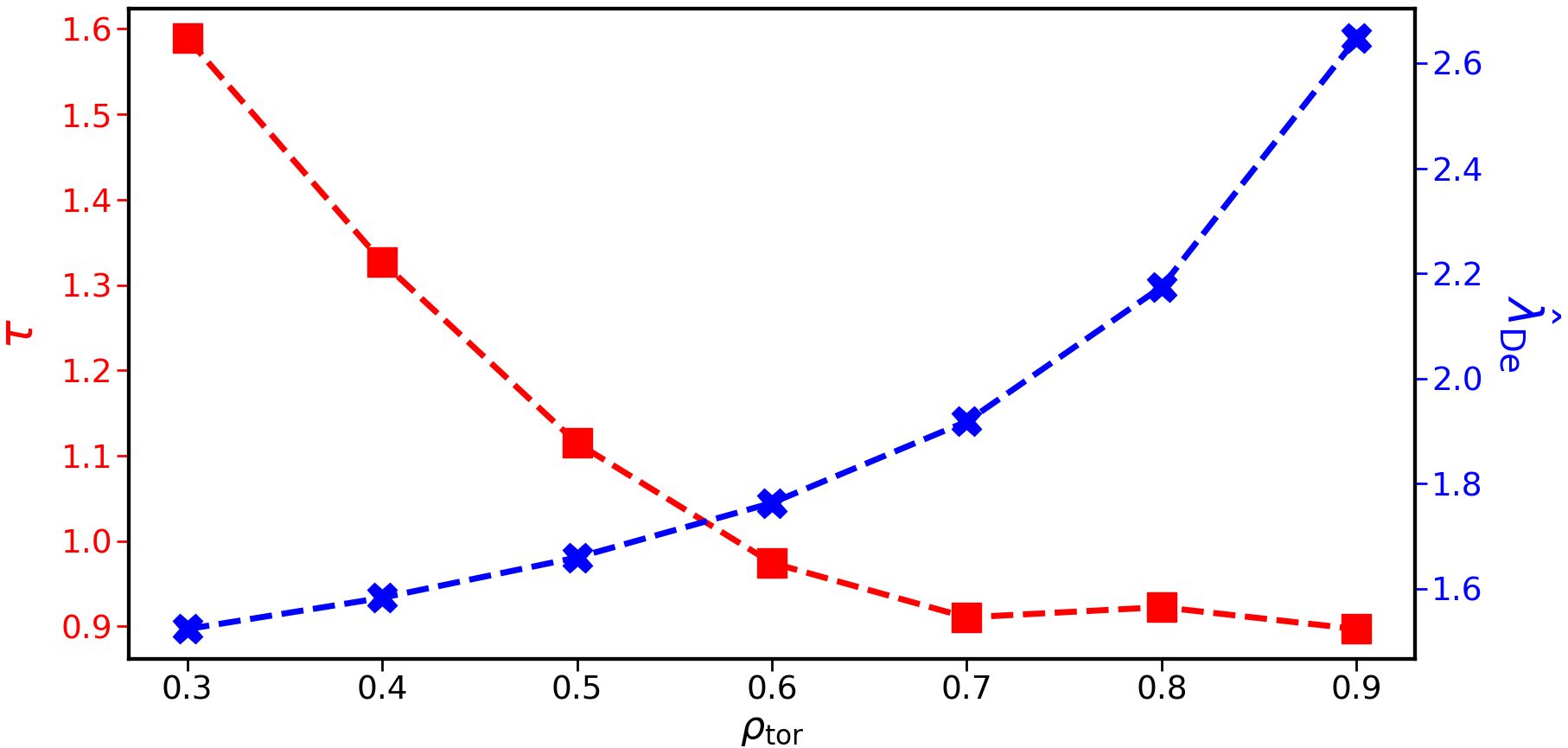}
    \caption{Variation of the temperature ratio $\tau=T_{\rm e}/T_{\rm i}$ (red) and the normalised Debye length (blue) as a function of the radial coordinate.}
    \label{fig:tau_vs_lambda}
\end{figure}

\begin{figure}[h!]
    \centering
    \includegraphics[width=\textwidth]{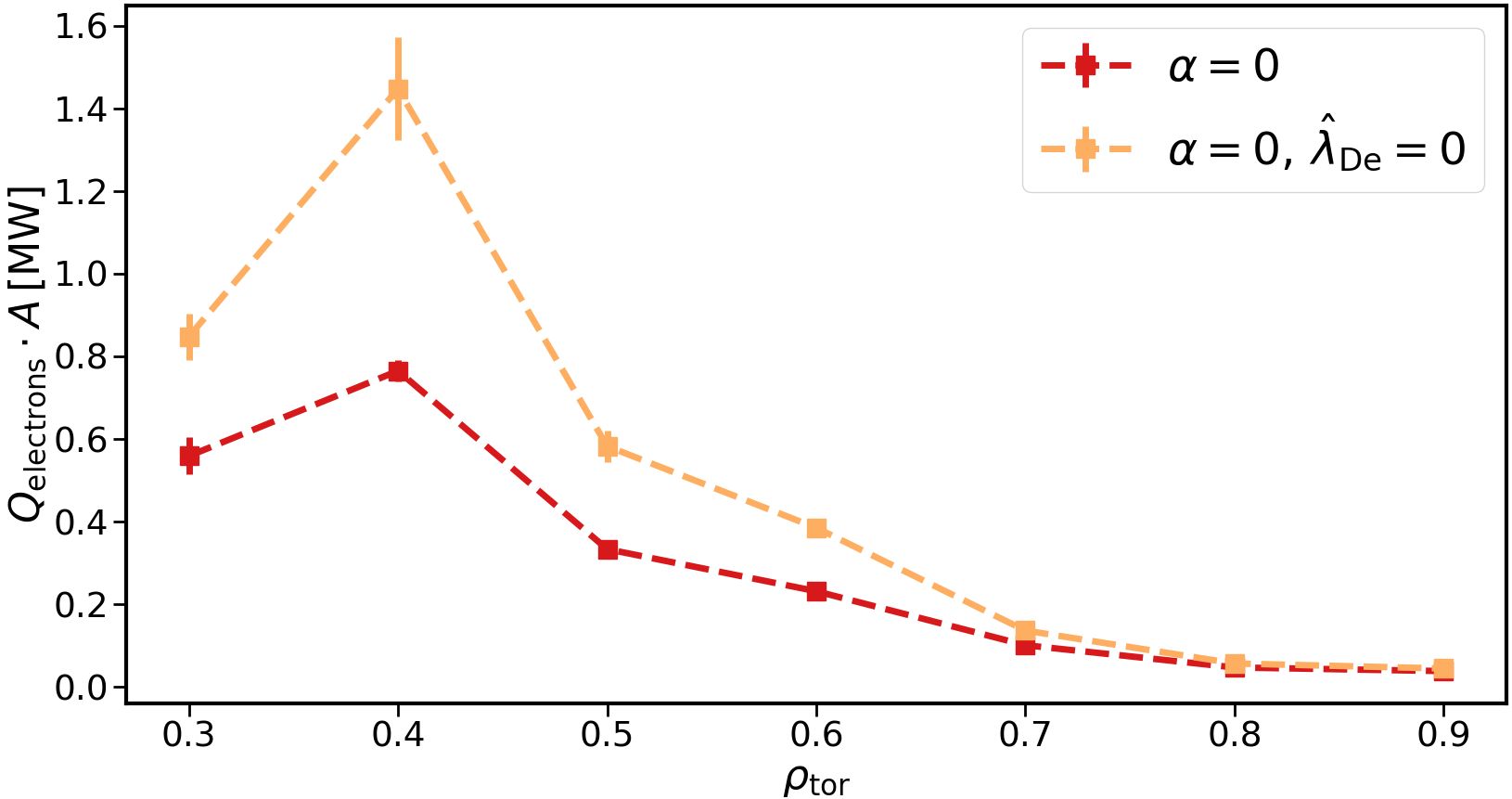}
    \caption{Radial electron-scale heat flux profiles of the $\alpha=0$-flux-tube with (red) and without (orange) including $\hat{\lambda}_{\rm De}$.}
    \label{fig:ETG_lambda}
\end{figure}

In order to understand the discrepancy between the electron-scale contributions at different radial positions, one has to consider several possible factors. As illustrated in figure \ref{fig:ETG_alpha}, the heat flux exhibits a near-constant behaviour despite variations in the field-line label. Hence, the selection of the flux-tube or subsequent averaging procedures should be negligible for the remaining discussion.
\noindent Previous investigations into ETG dynamics in Wendelstein 7-AS \cite{jenko2002stellarator} and similar simulations conducted in tokamaks \cite{gene} have suggested that collisionality and plasma-$\beta$ exert only minor influence on the electron-scale transport levels. However, the same studies indicated that the electron-to-ion temperature ratio $\tau=T_{\rm e}/T_{\rm i}$ and the normalised Debye length $\hat{\lambda}_{\rm De}=\lambda_{\rm De}/\rho_{\rm e}=\sqrt{B_{\rm ref}^2/(4\pi c^2 m_{\rm e} n_{\rm e}(x))}$ notably affect ETG turbulence. As depicted in figure \ref{fig:tau_vs_lambda}, the reduction in $\tau$ does not seem to account for the decreasing contribution of ETG to the overall flux, as q decrease in $\tau$ theoretically implies destabilisation toward the outer region.
Beyond the variation in $\tau$, figure \ref{fig:tau_vs_lambda} shows that $\hat{\lambda}_{\rm De}$ increases towards the edge, which is thought to have a stabilising influence on ETG turbulence. We repeated the flux-tube simulations while artificially setting $\hat{\lambda}_{\rm De}=0$ to investigate this. The results presented in figure \ref{fig:ETG_lambda} indicate that the stabilisation attributed to Debye shielding falls short in explaining the radial changes in electron-scale transport. Notably, the difference between both models diminishes significantly towards the edge.
In addition to geometric factors like magnetic field curvature, the normalised gradient ratio $\eta_{\rm e}=L_{\rm n}/L_{\rm T_{\rm e}}$ is known to influence transport significantly, as a decrease tends to stabilise ETG turbulence. This effect was proposed to explain the low electron-scale transport in the outer region of a discharge discussed in \cite{plunketg}. If this were the primary driver, neglecting the density gradient should logically yield a radial flux profile that increases towards the edge together with the normalised electron temperature gradient, which drives ETG.

\begin{figure}[h!]
    \centering
    \includegraphics[width=\textwidth]{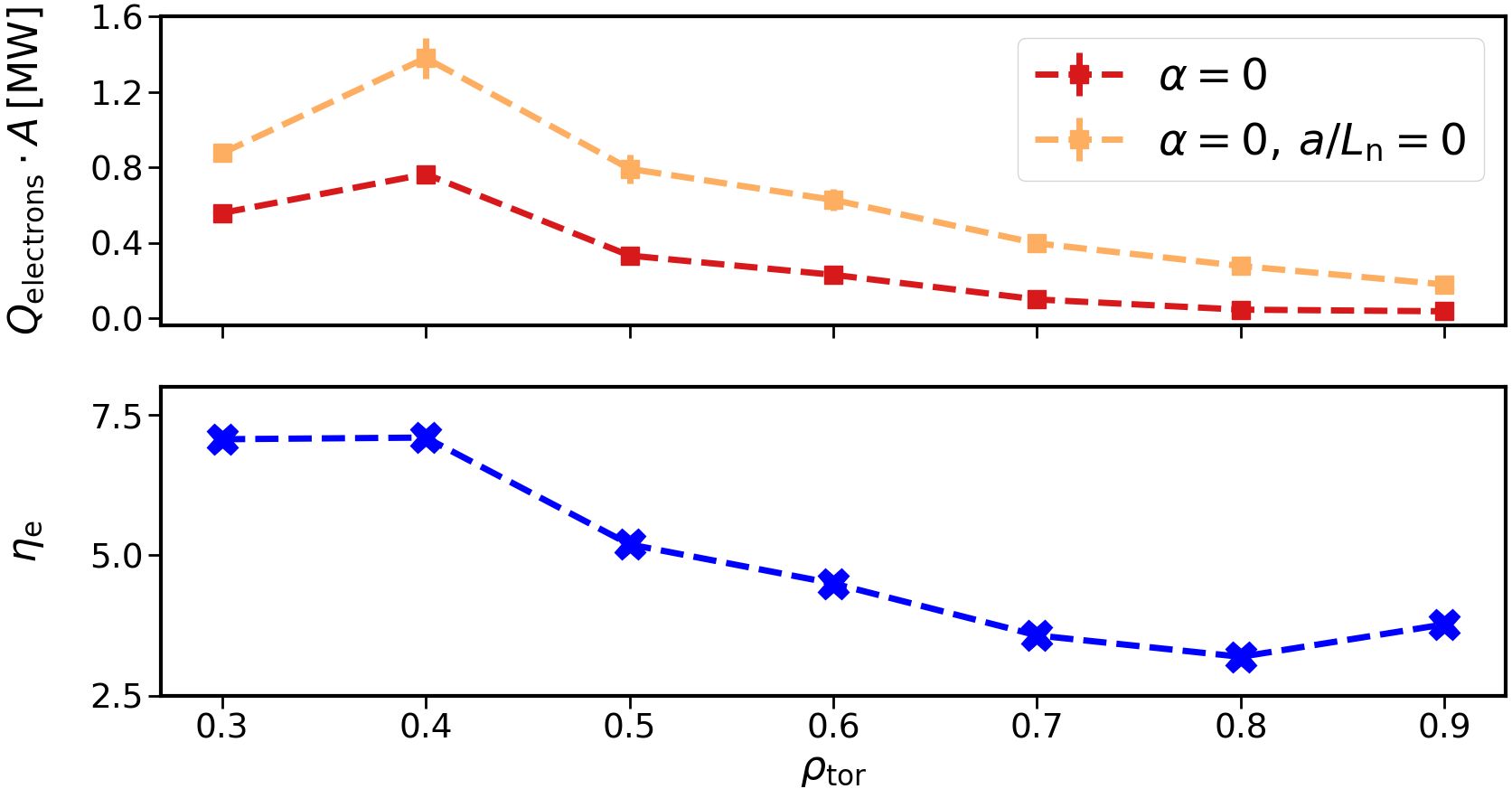}
    \caption{Top: Radial electron-scale heat flux profiles of the $\alpha=0$-flux-tube with finite $a/L_{\rm n}$ (red) and with $a/L_{\rm n}=0$ (orange); bottom: radial variation of the normalised gradient ratio $\eta_{\rm e}$.}
    \label{fig:ETG_eta}
\end{figure}

However, while the decrease of $\eta_{\rm e}$ along the radial direction aligns with the notion of stabilisation through the density gradient, figure \ref{fig:ETG_eta} reveals that eliminating the density gradient still leads to a decrease in heat flux towards the edge.

\begin{figure}[h!]
    \centering
    \includegraphics[width=\textwidth]{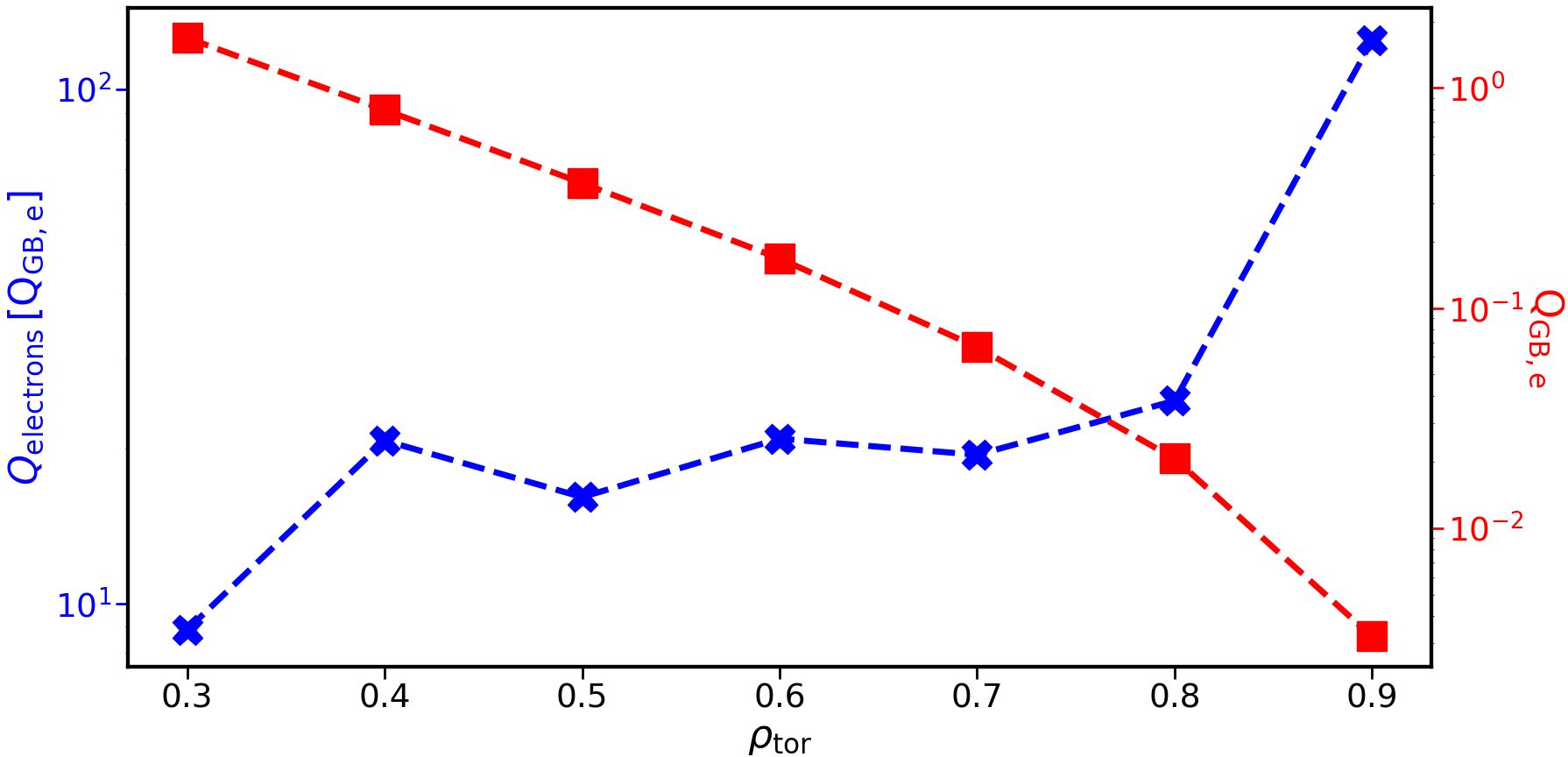}
    \caption{Blue: Radial electron-scale heat flux profile in Gyrobohm units, averaged over all field lines considered, using nominal simulation parameters; red: radial variation of Gyrobohm transport. The subscript 'e' indicates that the electron mass was used as reference mass $m_{\rm ref}$.}
    \label{fig:ETG_GB}
\end{figure}

Having eliminated all other options, we conclude that ETG transport is not stiff enough to compensate for the decrease in density and temperature towards the outer regions. To confirm this, we show the radial profile of the normalised electron heat flux and the Gyrobohm scaling factor profile in figure \ref{fig:ETG_GB}.
\noindent In there, we see that the normalised heat flux notably increases by more than an order of magnitude towards the edge, coinciding with the growing normalised electron temperature gradient. However, the Gyrobohm scaling factor drops rapidly over two orders of magnitude from the core to the edge. Consequently, it becomes evident that the normalised flux fails to grow fast enough with the background drive. As a result, the product of these two factors decreases as the radial position increases.

\section{Assessment of the particle flux}
\label{sec:power_balance_comments}
While we have presented a first-of-its-kind simulation of experimentally relevant parameters of a gas-fuelled ECRH discharge in W7-X, it is reasonable to question the validity of the conclusions drawn. This scepticism is justified as the simulated heat fluxes still fail to agree with the power balance, even with the additional electron-scale contributions. Particularly, establishing a correlation between simulation and experiment for the outer region beyond $\rho_{\rm tor}=0.5$ is challenging due to the overestimation observed in the ion heat flux calculated by GENE-3D. A more robust approach would involve a comprehensive coupling of several codes, such as GENE-3D, GENE (for ETG simulations), a neoclassical solver like KNOSOS \cite{velasco2020knosos}, and other codes, to a transport code like Tango \cite{parker2018bringing,parker2018investigation,shestakov2003self}, in order to iteratively evolve the background profiles self-consistently until agreement with power balance is reached. While such an approach is routinely done for tokamaks \cite{di2022global,di2023predictions} already, it is only done globally using an adiabatic electron model \cite{navarro2023first} or within a flux-tube framework with kinetic electrons \cite{mandell2023stellarator} for stellarator geometries. Expanding the GENE-3D-Tango methodology to incorporate kinetic electrons currently exceeds the scope of this paper but remains a target for the future.

Nevertheless, we can further strengthen the argument for the importance of TEM and ETG turbulence in the core of gas-fuelled ECRH plasmas without matching power balance by examining the individual contributions to the electron particle flux. While the experimental flux value remains unknown due to significant systematic uncertainties, it is plausible to assume that, for a purely gas-fuelled discharge, the total particle flux will tend towards zero or, at least, be very close to it deep within the plasma core.
\begin{figure}[h!]
    \centering
    \includegraphics[width=\textwidth]{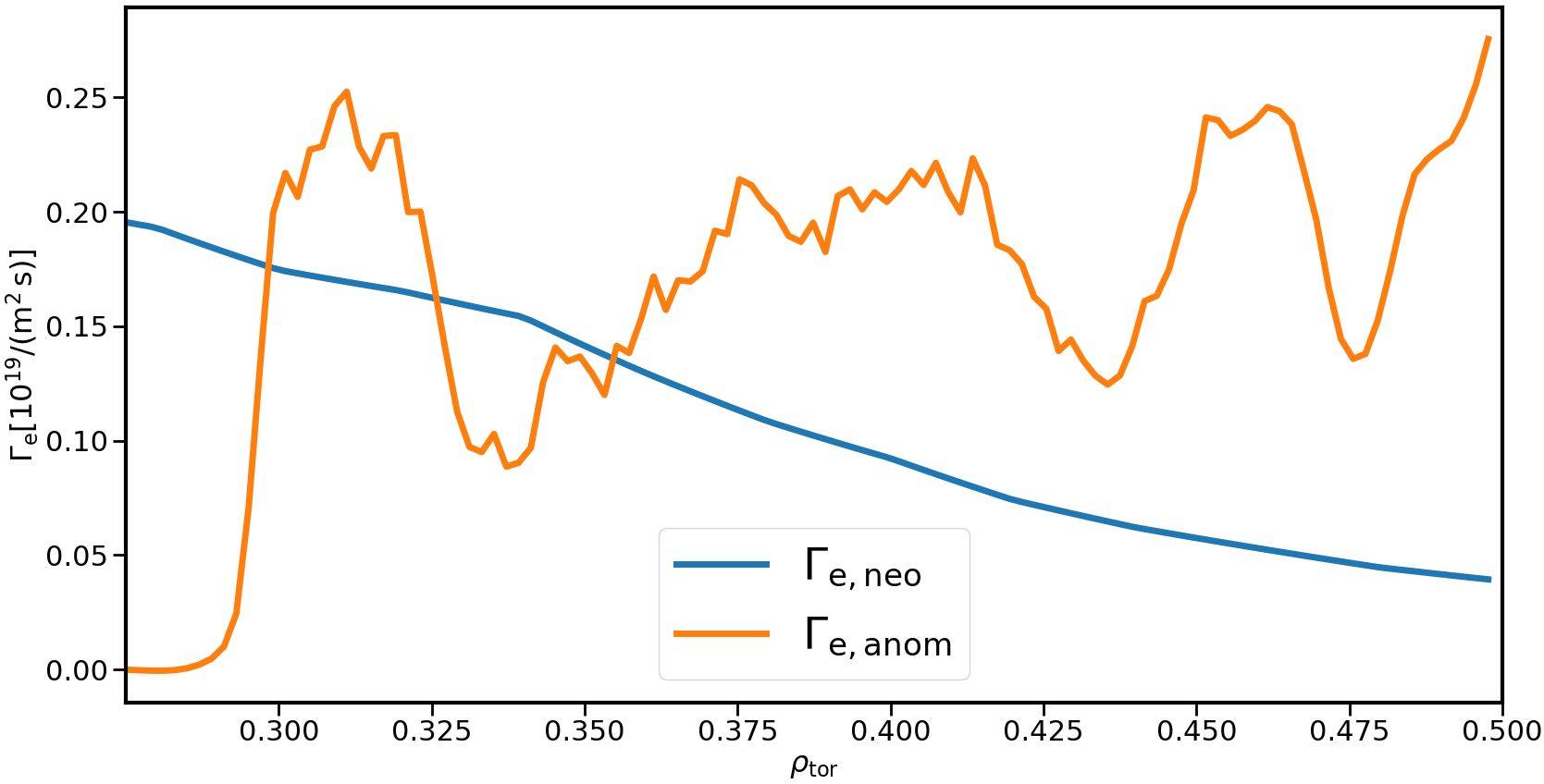}
    \caption{Radial profiles of contributions to the electron particle flux. The neoclassical particle flux, calculated with DKES, is shown in blue. In contrast, the anomalous flux, calculated with GENE-3D, is shown in orange.}
    \label{fig:particle_flux_core}
\end{figure}
\noindent A qualitative assessment of the particle flux levels depicted in figure \ref{fig:particle_flux_core}, when compared with findings from \cite{thienpondt2023prevention}, suggests that the anomalous particle flux predicted by GENE-3D appears too positive to align with the experiment. Typically, neoclassical transport is directed outward\cite{maassberg1999density,beidler2018expected}, necessitating an inward turbulent particle flux when approaching the magnetic axis, a trend not observed here. Additionally, as highlighted in \cite{thienpondt2023prevention}, an inward anomalous particle flux can be achieved by reducing the density gradient, increasing the electron temperature gradient, or a combination of both. Considering that these adjustments would further increase the drive for $\nabla T_{\rm e}$-TEM and ETG turbulence, it stands to reason that when matching experimental heat and particle fluxes, the contributions of electron-induced turbulence to the electron heat flux would likely be even more pronounced than under the nominal profiles used in this analysis.

In summary, our findings provide robust evidence for the presence of trapped-electron turbulence within the core of gas-fuelled ECRH discharges, manifested in the form of ITG-TEM hybrids and ETG turbulence. Their relevance in high-performance discharges, such as those employing pellet fuelling, remains an avenue yet to be explored.

\section{Conclusions}
This paper delves into plasma turbulence analysis in an experimental ECRH discharge of Wendelstein 7-X using GENE-3D and GENE. Our study unveils that while flux-tube and full-flux-surface simulations reasonably approximate the heat flux levels predicted by radially global simulations within the plasma core, they significantly overestimate transport in the outer plasma region. However, factoring in the shearing of external ExB-flow helps mitigate this discrepancy to some extent, underscoring its pivotal role in stabilising turbulence, even in standard stellarator discharges.

Moreover, we have provided substantial evidence supporting the existence of trapped-electron-mode turbulence within the plasma core, challenging previous notions proposed in existing literature \cite{klinger,grulkeitgcore}. Although not solely accountable for transport, these modes were distinctly evident in the form of ITG-TEM hybrids, predominantly driven by the electron temperature gradient.

Additionally, our study demonstrates, within our specific case, the substantial contribution of electron-scale simulations to electron heat flux in contrast to the suggestion of weak ETGs in W7-X outlined in \cite{plunketg}. We observed a diminishing impact of these modes in the plasma's outer radial region, attributed to the normalised ETG transport not increasing rapidly enough towards the edge to offset the decrease in the Gyrobohm scaling factor resulting from lower electron temperature and density.

Finally, we support the validity of our findings despite not aligning with transport predictions from power balance analysis. We assert that the impact of electron-induced turbulence is expected to significantly increase if background profiles are modulated to achieve experimentally realistic heat and particle fluxes.

In the future, we will extend these studies to a broader range of experimental discharge in order to generalise our claims, while additionally coupling GENE-3D simulations to the transport code Tango as shown in \cite{navarro2023first,navarro2023assessing} in order to achieve flux matching with the experiment. In addition, global simulations offer a direct comparison with turbulent fluctuation measurements via synthetic diagnostics \cite{hansen2022development} in order to identify signals of trapped electron modes in experimental discharges.

\section*{Acknowledgments}
This work has been carried out within the framework of the EUROfusion Consortium, funded by the European Union via the Euratom Research and Training Programme (Grant Agreement No 101052200 — EURO-fusion). Views and opinions expressed are however those of the author(s) only and do not necessarily reflect those of the European Union or the European Commission. Neither the European Union nor the European Commission can be held responsible for them. Numerical simulations were performed at the Cobra and Raven HPC system at the Max Planck Computing and Data Facility (MPCDF), Germany and the Marconi Fusion supercomputer at CINECA, Italy.
\printbibliography

\end{document}